\documentclass[%
 aip,
 amsmath,amssymb,
 reprint,%
]{revtex4-1}
\usepackage{graphicx}
\usepackage{dcolumn}
\usepackage{bm}
\usepackage{multirow}
\usepackage{enumerate}
\usepackage{color}
\usepackage{subcaption}
\usepackage[utf8]{inputenc}
\usepackage[T1]{fontenc}
\usepackage{mathptmx}
\usepackage{etoolbox}
\newcommand{\ds}{\displaystyle}
\newtheorem{theorem}{Theorem}[section]

\newtheorem{remark}[theorem]{Remark}

\newtheorem{proposition}[theorem]{Proposition}

\newtheorem{example}[theorem]{Example}

\makeatletter
\def\@email#1#2{%
 \endgroup
 \patchcmd{\titleblock@produce}
  {\frontmatter@RRAPformat}
  {\frontmatter@RRAPformat{\produce@RRAP{*#1\href{mailto:#2}{#2}}}\frontmatter@RRAPformat}
  {}{}
}%
\makeatother
\begin{document}

\preprint{AIP/123-QED}

\title{Increasing the Synchronization Stability
in Complex Networks}
\author{Xian Wu}
\author{Kaihua Xi}%

 \thanks{Corresponding author, email: kxi@sdu.edu.cn}
 \author{Aijie Cheng}
\affiliation{ 
School of Mathematics, Shandong University, Jinan, Shandong, 250100, China
}%

\author{Hai Xiang Lin}
\author{Jan H. van Schuppen}
\affiliation{%
Delft Institute of Applied Mathematics, Delft University of Technology, Delft, 2628 CD, The Netherlands
}%

\date{\today}

\begin{abstract}
We aim to increase the ability of a of coupled phase oscillators to maintain the synchronization when 
the system is affected by stochastic disturbances. 
We model the disturbances by Gaussian noise and use 
the mean first hitting time when the state hits the boundary of a 
secure domain, that is a subset of the basin of the attraction, to measure the synchronization stability. 
Based on the invariant probability distribution of a system of phase oscillators subject to Gaussian
disturbances, we propose an optimization method to increase the mean first hitting time, and thus
increase the synchronization stability. In this method, 
a new metric for the synchronization stability is defined as the probability of the state being absent from the secure domain, which reflects the impact of all the system parameters and the strength of the disturbances. Furthermore, by this new metric, one may identify those edges which may lead to desynchronization with a high risk.  
A case study shows that the mean first hitting time is dramatically increased after solving 
the corresponding optimization problems and the vulnerable edges are effectively identified.  It is also found that optimizing the synchronization by maximizing the order parameter or the phase cohesiveness may dramatically increase the value of the metric and decrease the mean first hitting time, thus decrease the synchronization stability. 
\end{abstract}

\maketitle

%

\section{Introduction}

Synchronization of coupled phase oscillators has served as a paradigm for understanding collective 
behavior of real complex systems, where examples arise in nature (e.g., chimera spatiotemporal patterns \cite{Abrams}, cardiac pacemaker cells\cite{cardiac}) 
and artificial systems (e.g., multi-agent systems\cite{macuiqin}, distributed optimization\cite{YANG2019278}, power grids \cite{Xi2016,motter}). 
For systems such as a power grid, if the synchronization is lost, then the system can no longer function properly.
The objective of this paper is to propose a method to increase the ability of these systems 
to maintain the synchronization under disturbances, which is 
called the synchronization stability. 
We introduce a new metric for the analysis of the synchronization stability, which 
does not only reflect the role of the system parameters, i.e., the natural frequency, the network topology 
and the coupling strength but also reflect the role of the strength of the disturbances at the nodes. 
With this metric as the objective of an optimization framework, 
the synchronization stability can be optimized by redistributing either the natural frequencies of nodes
and the coupling strength of edges. In addition, the vulnerable edges that limit 
the synchronization stability can be effectively identified by this metric.  The result of this paper provides a new avenue for the analysis of the synchronization stability 
of the complex networks. 

On the synchronization of the complex networks, significant insights have been obtained from investigations on the emergence of a synchronous state and synchronization coherence.
The synchronization is determined by the system parameters, including the natural frequencies at the nodes, the network topology and the coupling strength of edges. With the metrics of
the critical coupling strength \cite{Dorfler20141539,FAZLYAB2017181} and the order parameter\cite{Skardal2014}, the influences of these parameters on the synchronization are widely investigated. Based on these investigations the system parameters may be assigned to optimize the synchrony, which can be attained by deletion or addition of edges, or by changing the coupling strength of the edges in the network. An important problem is to maintain the synchronization when the system is subjected to disturbances. Regarding the ability to maintain the synchronization, the spectrum of the system matrix of the linearized system and the volume of the basin of attraction of a stable synchronous state may be investigated \cite{pecora,menck,sizeofbasin} . However, in these investigation, the severity of the disturbances are not considered and the edges at which the synchronization may be lost cannot be effectively identified.
%
%
\par 
In control theory, the synchronous state is also mentioned as the \emph{set point} for control, in which control actions are taken to let the state converge to the synchronous state after disturbances. Thus, with frequently occurring disturbances, the phase may fluctuate around the synchronous state. 
If the fluctuations in the phase differences are so large that the state of the system cannot stay inside a neighbourhood of the synchronous state, then the synchronization is lost. We say an edge is more vulnerable if the desynchronization occurs at this edge more easily. 
The $\mathcal{H}_2$ norm of a linear input-output system is often used to study the synchronization performance after the disturbances \cite{optimal_inertia_placement,H2norm,RobustnessSynchrony}. By minimizing this $\mathcal{H}_2$ norm 
as an objective and the system parameters as decision variables, the fluctuations 
in the phase differences may be effectively suppressed. In a framework of the theory of 
stochastic processes, the dependence (or relationship) between the fluctuation of the phase difference in each edge and the system parameters is revealed, in which the cycle space of graphs plays a role \cite{KaihuaXi2022}.  
\par 
However, it is insufficient to focus on the fluctuations in the phase differences only for the synchronization stability analysis. In fact, the risk of losing synchronization is actually determined by two factors, i.e., the fluctuations of the state and the size of the 
basin attraction of the synchronous state. 
Note that due to the nonlinearity of the system, the fluctuations 
of the state also depend on the synchronous state\cite{KaihuaXi2022}.
Thus, to increase the synchronization stability of a system with disturbances, it is important to
find such a synchronous state that has a large basin of attraction and around which 
the fluctuation of the state is also small.  The concept of the first hitting time of a stochastic process, which is a random variable, is often used to study the stability of nonlinear systems \cite{stochasticstability, survivalanalysis}. 
For the stability analysis of the coupled phase oscillators, the first hitting time 
can be defined as the first time when the state starting at the synchronous state hits the boundary of the basin of attraction. Clearly, this first hitting time depends on both the size of the basin of attraction of the synchronous state and the fluctuations of the 
state.  The larger the mean of this first hitting time, the higher probability of the state staying in the basin of attraction and 
the stronger ability to maintain the synchronization.  However, due to the nonlinearity and high dimension of the system, the boundary of the basin of attraction of the synchronous state can hardly be precisely 
estimated. A sign of losing the synchronization is that there are edges in which the 
absolute values of the phase differences become larger than $\pi/2$ and then go to infinity as time increases
to infinity.
Thus, we focus on the domain in which the absolute values of the phase differences
in the edges are all smaller than $\pi/2$, which is called \emph{the secure domain}
in this paper. Clearly, 
if the phase differences in the edges are in the secure domain for all the time, 
the system maintains the synchronization \cite{Kuramotostability}. Once the state goes out of this secure domain, 
the synchronization may be lost. Hence, with this secure domain, the concept of the first hitting time can be applied to the complex system.
\par 
In this paper, we model  the frequently occurring disturbances in the nonlinear dynamics
by Gaussian noise and investigate the risk of the state going out of the secure domain in the corresponding nonlinear stochastic process. If one linearizes the nonlinear stochastic system then the resulting linear stochastic system driven
by a Brownian motion process has a Gaussian invariant probability distribution.
Based on this invariant probability distribution, we define a metric for the 
risk of the state of the nonlinear stochastic process going out of the secure domain and propose
an optimization framework to minimize this metric, thus increase the mean first time when 
the state starting at the synchronous state hits the boundary of the secure domain. We show 
the range of this metric and by the optimization framework, we address the design problem of the coupling strength and the natural frequency respectively. It will be shown that after 
maximizing the probability of the states of the Gaussian process inside the secure domain, the mean first hitting 
time is effectively increased, which indicates an increase of synchronization stability. 
\par
This paper is organized as follows. The model of the coupled phase oscillators is introduced in Section \ref{sec:introduction}. We describe the concept of the mean first hitting time and 
the invariant probability distribution of the linear stochastic process in Section \ref{sec:firsthittingtime} and \ref{Sec:invariantprobability} and 
propose an optimization method to decrease the risk of the state being absent from the secure domain in 
Section \ref{sec:optimization}. A case study for the evaluation of the performance of 
the optimization framework is presented in Section \ref{sec:casestudy}.
We conclude this paper with perspectives in Section \ref{sec:conclusion}.

\section{The model}\label{sec:introduction}
We consider an undirected graph $\mathcal{G}=(\mathcal{V},\mathcal{E})$ with $n$ nodes in the set $\mathcal{V}$ and $m$ edges in the set $\mathcal{E}$. The dynamics of the coupled phase oscillators are described by the following differential equations, 
\begin{equation}
  \label{Kuramoto Equation}
  \dot{\varphi_i}(t)=\omega_i-\sum\limits_{j=1}^n l_{ij}\sin(\varphi_i(t)-\varphi_j(t)),\, \text{for}\quad i=1,2,\cdots,n,
\end{equation}
where $\varphi_i$ is the phase of oscillator $i$, $\omega_i$ represents the natural frequency, $l_{ij}$ denotes the coupling strength of the edge $(i,j)\in\mathcal{E}$ which connects nodes $i$ and $j$ and $l_{ij}>0$ if nodes $i$ and $j$ are connected and $l_{ij}=0$ otherwise. It is assumed that the graph is connected, thus it holds $m\geq n-1$. 
\par
Without loss of generality, we assume 
that $\sum_{i=1}^n{\omega}_i=0$ and there exists a synchronous state $\bm\varphi^*=\text{col}(\varphi_i)\in\mathbb{R}^n$ such that 
\begin{equation}
    \label{synchronous state}
  \omega_i-\sum\limits_{j=1}^n l_{ij}\sin(\varphi_i^{*}-\varphi_j^{*})=0,\,~~i=1,2,\cdots,n,
\end{equation}
which can be typically obtained by increasing the coupling strength of the edges. 
We focus on the synchronous state in the following domain, 
\begin{equation}
    \label{security condition}
    \Theta =\{\bm\varphi\in\mathbb{R}^{n}\big||\varphi_{i}-\varphi_{j}|<\pi/2,\forall (i,j)\in\mathcal{E}\}.
\end{equation}
which in this paper is called \emph{the secure domain} for the stability analysis. It has been shown 
that the synchronous state in this domain is asymptotically stable and by the Lyapunov method for 
stability analysis, the state of system (\ref{Kuramoto Equation}) starting inside this domain will converge to a synchronous state in this domain\cite{Kuramotostability}. 
If the synchronization is lost, the state of the system must have gone out of this secure domain. Conversely,
if the state of the system stays in this domain at any time, the synchronization is maintained.
Thus,  to increase the synchronization stability, it is critical to decrease the risk that 
the state leaving this secure domain. 
\par
Due to the disturbances brought to the natural frequency, the system may lose its synchronization. 
The application of a perturbation is an effective way
to study the fluctuations of the state caused by the disturbances, in which the focus is the dynamics
\begin{align}\label{deterministicmodel}
 \dot{\varphi_i}(t)=\omega_i-\sum\limits_{j=1}^n l_{ij}\sin(\varphi_i(t)-\varphi_j(t))+\Delta \omega_i(t),~~ i=1,\cdots,n
\end{align}
where $\Delta\omega_i(t)$ denotes the frequently occurring disturbance at node $i$.
In this paper, the theory of stochastic process is used to study the synchronization stability. 
We model the disturbance $\varphi \omega_i$ by Gaussian noise and focus on the following stochastic process,
\begin{equation}
  \label{Stochastic Equation}
  \dot{\varphi_i}(t)=\omega_i-\sum\limits_{j=1}^n l_{ij}\sin(\varphi_i(t)-\varphi_j(t))+b_i w_i(t),\,  \text{for}~ i=1,\cdots,n.
\end{equation}
where the variable $w_i(t)$ represents a standard Gaussian white noise process affecting 
node $i$.
For any two distinct nodes $i$ and $j$, the stochastic processes $w_i$ and 
$w_j$ are assumed to be independent. The variable $b_i$ specifies 
the standard deviation of the noise. It is remarked that equation (\ref{deterministicmodel}) describes a deterministic system while equation (\ref{Stochastic Equation}) describes a stochastic system. The latter system models in more detail the fluctuations of the state of the system
due to disturbances and hence is more suitable to investigate the synchronization stability in case of
such disturbances. In addition, the disturbance in (\ref{deterministicmodel}) may be bounded 
while the one modeled by the Gaussian noise in (\ref{Stochastic Equation}) is unbounded.
\par 
When the system loses its synchronization, there is at least one edge in which the absolute value
of the phase differences goes to infinity as time increases to infinity. 
We denote $e_k=(i,j)\in\mathcal{E}$ for $k=1,\cdots,m$.
To obtain information about the phase differences of all the edges in the network, we define the output of the system (\ref{Stochastic Equation}) as those phase difference according to the formula, 
\begin{equation}\label{output-linear}
      y_k(t)=\varphi_i(t)-\varphi_j(t), ~\text{for}~ k=1,\cdots,m,
\end{equation}
where $k$ is the index of edge $e_k=(i,j)$ in  the edge set $\mathcal{E}$. 
Here, the direction of edge $e_k$ is from node $i$ to $j$, which is required to obtain the 
phase differences in the output. This direction is arbitrarily specified, which has no impact on the following analysis.  In the remainder of this paper, the vector notations $\bm \varphi(t)=\text{col}(\varphi_i(t))\in\mathbb{R}^n$ for the state variables in (\ref{Stochastic Equation}) and $\bm y(t)=\text{col}(y_k(t))\in\mathbb{R}^m$ for the output in (\ref{output-linear}) will be used for simplicity.  Corresponding to the phase $\bm \varphi^*$ at the synchronous state, the output is denoted by 
\begin{align}
\bm y^*&=\text{col}(y_k^*)\in\mathbb{R}^m~\text{with}~y_k^*=\varphi_i^*-\varphi_j^*. \label{expectation}
\end{align}
\par 
For the deterministic system (\ref{Kuramoto Equation}), with the metrics of \emph{the order parameter}, or \emph{the critical coupling strength}, or 
\emph{the phase cohesiveness} that is the $\mathcal{L}_\infty$ norm of $\bm y^*$, the synchronization may be improved by designing the network topology and redistributing the natural frequencies. See Appendix \ref{optimization-Case-study} 
for the definition of the order parameter. Here,
the critical coupling strength is defined as the smallest coupling strength 
of the edges at which a phase transition from incoherency to synchronization occurs \cite{DorflerCriticalcoupling}.
 
\par 

\section{The mean first hitting time}\label{sec:firsthittingtime} 
The first hitting time model is often used to study the survival time of a system\cite{survivalanalysis}, which is also 
used to study the stability of nonlinear systems\cite{stochasticstability}.   In a first hitting time model, 
there are two components, i.e., a stochastic process $\{\bm x(t)\in \mathbb{X},  t\in \mathbb{T}\}$ with initial 
value $\bm x(0)=\bm x_0$,  where $\mathbb{X}$ is the state space of the process, a boundary set 
$\mathbb{B}\subset\mathbb{X}$ and $\mathbb{T}=[0,+\infty)$. Assume that the initial value of the process $x_0$ lies outside 
of the boundary set $\mathbb{B}$, then the first hitting time can be defined by the random variable $t_e:\Omega\rightarrow\mathbb{T}$, 
\begin{align}
t_e=
\begin{cases}
\inf_{t\in \mathbb{T}} x(t)\in \mathbb{B}, &\text{if such a } t\in\mathbb{R}_+~~\text{exists},\\
+\infty,  &\text{else},
\end{cases}
\end{align}
where $t_e$ is the first time when the sample path of the stochastic process reaches the boundary set $\mathbb{B}$.
The first hitting time is also called \emph{the first exit time} when the sample path of the stochastic process exits a set $\mathbb{A}$ with  $\partial\mathbb{A}=\mathbb{B}$ and the initial state lying inside $\mathbb{A}$. 
Clearly, this first hitting time depends on the probability distribution function of the stochastic process $\bm x(t)$, the initial value 
and the boundary set $\mathbb{B}$.  For some specific stochastic processes, such as the Wiener process and the Ornstein-Uhlenbeck process, the probability density of the first hitting time can be analytically derived 
\cite{ALBEVERIO1997139,passagetimeLuigi-1988}. For a complex 
stochastic process such as the one described by (\ref{Stochastic Equation}), the moment of the first hitting time can be approximated by the Monte Carlo method, i.e., given an initial value and a boundary set, the distribution of the first hitting time can be approximated by simulating the stochastic process, then the moment can be computed with a large amount of the simulations. 
\par 
For the system (\ref{Stochastic Equation}), to use the 
first hitting time model, the boundary set can be $\mathbb{B}=\partial\mathbb{A}$ where the set $\mathbb{A}$ denotes 
the basin of the attraction and the state space $\mathbb{X}=\mathbb{R}^n$. 
Clearly,  similar as the synchronization stability, the expectation of the first hitting time depends on the size of the basin of the attraction and the severity of the disturbances. 
Thus, the expectation of the first time when the state hits the boundary of the basin of the attraction can be used to characterize the synchronization stability. However, due to the difficulty in estimating the boundary of the basin of attraction, the expectation of the first exit time is difficult to be precisely estimated even by statistics of simulations based on the Monte Carlo method.
Alternatively, the first exit time of the state from the secure domain $\Theta$, rather than from  the basin of the attraction, is used to characterize the synchronization stability. Correspondingly, in the first hitting time
model, one chooses the boundary of the secure domain according to $\mathbb{B}=\partial\Theta$
and $\mathbb{A}=\Theta$. A larger first hitting time implies a longer period of synchronization stability and an increased
stability against disturbances. Because this secure domain is a subset of the basin of attractions of the synchronous state, this first hitting time is smaller than the survival time of the system.
\par 
The distribution of the first hitting time is closely related to the probability density function of the state of system (\ref{Stochastic Equation}). 
The probability density of the system (\ref{Stochastic Equation}) can be solved from the corresponding \emph{Forward Kolmogorov Equation}\cite{stochasticstability},  which however is very complex because of the high dimension of the system.   Thus, we do not aim to derive the analytical form of the probability density function of the first hitting time but focus on its mean $\overline{t}_e$ which is computed approximately by the Monte Carlo method in which a large amount of  simulations of (\ref{Stochastic Equation}) with initial state $\bm x_0$ at the synchronous state are performed.
These simulations will be carried out in the section of case study to show the changes in the synchronization stability.
This is practical because the stochastic disturbances, which may be independent on the state, occur continuously even when the state is at the synchronous state. 
\par 
\section{The invariant probability distribution of a linear stochastic process} \label{Sec:invariantprobability} 
Now, we focus on the following linear stochastic system,
\begin{equation}
    \begin{aligned}
      \label{linear System}
      \dot{\widehat{\bm\varphi}}(t)&=-\bm L_a\widehat{\bm\varphi}(t)+\bm B\bm w(t),\\
      \widehat{\bm y}(t)&=\bm C^{\top}\widehat{\bm\varphi}(t),
    \end{aligned}
\end{equation}
which is linearized from (\ref{Stochastic Equation}) at the synchronous state $\bm \varphi^*$. 
Here, the state variable $\widehat{\bm \varphi}$ and the output $\widehat{\bm y}(t)$ represent the deviation of the state $\bm\varphi(t)$ from $\bm \varphi^*$ and of the output $\bm y(t)$ from $\bm y^*$ 
respectively, 
 $\bm L_a=(l_{a_{ij}})
 \in\mathbb{R}^{n\times n}$ is the Laplacian matrix such that 
\begin{equation*}
    l_{a_{i j}}= 
    \begin{cases}
       -l_{ij}\cos(\varphi_i^{*}-\varphi_j^{*}), & i \neq j, \\
       \sum\limits_{k\neq i}l_{ik}\cos(\varphi_i^{*}-\varphi_k^{*}), & i=j,
      \end{cases}
\end{equation*}
$\bm B=\text{diag}(b_i)\in\mathbb{R}^{n\times n}$ is a diagonal matrix, $\bm w=\text{col}(w_i)\in\mathbb{R}^n$ is 
Gaussian white noise and
$\bm C=(C_{ik})\in\mathbb{R}^{n\times m}$ is 
the incidence matrix of the graph $\mathcal{G}$ such that
\begin{equation}\label{incidence}
    C_{{i k}}= 
    \begin{cases}
       1,\, &\text{node $i$ is the begin of  edge $e_k$,} \\
       -1,\, &\text{node $i$ is the end of  edge $e_k$,} \\
       0,\, &\text{otherwise},
      \end{cases}
\end{equation} 
where the direction of the edge $e_k$ is specified as in the definition of $y_k$ in (\ref{output-linear}). 
Because $\bm L_a$ is symmetric and non-negative definite, there exists an orthogonal matrix $\bm U\in\mathbb{R}^{n\times n}$ such that 
\begin{equation}\label{spectral}
    \bm U^{\top}\bm L_a\bm U=\bm\Lambda,
\end{equation}
where $\bm\Lambda=\text{diag}(\lambda_i)\in\mathbb{R}^{n\times n}$ with $0=\lambda_1\leq \lambda_2\leq\cdots\leq\lambda_n$ being the eigenvalues of matrix $\bm L_a$. The orthogonal matrix $\bm U$ can be written as $\bm U=[\bm u_1,\bm U_2]$, where $\bm u_1=\eta\bm 1_n,  \eta$ is a constant and $\bm U_2=[\bm u_2,\cdots,\bm u_m]\in\mathbb{R}^{n\times(n-1)}$, with the $i$-th column $\bm u_i$
of $\bm U$ being the eigenvector corresponding to the eigenvalue $\lambda_i$  for $i=2,\cdots,n$.
\par 
Due to the Gaussian distribution of $\bm w$, the state $\widehat{\bm \varphi}$ and the output $\widehat{\bm y}$ are also Gaussian, i.e.,
\begin{align*}
\widehat{\bm \varphi}(t)\in G\big(\bm m_{\widehat{\varphi}}(t),\bm Q_{\widehat{\varphi}}(t)\big),~~\widehat{\bm y}(t)\in G\big(\bm m_y(t),\bm Q_{\widehat{y}}(t)\big),
\end{align*}
with $\bm m_{\widehat{\varphi}}(t)\in\mathbb{R}^n,\bm Q_{\widehat{\varphi}}(t)\in\mathbb{R}^{n\times n}$ and $\bm m_{\widehat{y}}(t)\in\mathbb{R}^m,\bm Q_{\widehat{y}}(t)\in\mathbb{R}^{m\times m}$. 
Because the system matrix in (\ref{linear System}) is singular, the invariant probability distribution of $\widehat{\bm \varphi}(t)$ does not exist.  See Appendix \ref{InvariantDistribution} for the invariant distribution of a linear stochastic system. In order to obtain the invariant probability of the output $\widehat{\bm y}(t)$, we make the following transformation.
Let $\bm x(t)=\bm U^\top \widehat{\bm\varphi}(t)$. With the spectral decomposition of $\bm L_a$ in (\ref{spectral}),  we obtain
\begin{align}\label{transformed}
\dot{\bm x}(t)=-\bm \Lambda \bm x(t)+\bm U^\top \bm B \bm w(t).
\end{align}
Decompose the state $\bm x(t)$ and the matrix $\bm \Lambda$ into block matrices,
\begin{align*}
\bm x(t)=
\begin{bmatrix}
x_1(t)\\
\bm x_2(t)
\end{bmatrix},~
\bm\Lambda=
\begin{bmatrix}
0&\bm 0\\
\bm 0&\bm \Lambda_{n-1}
\end{bmatrix}\in\mathbb{R}^{n\times n}.
\end{align*}
where $\bm\Lambda_{n-1}\in\mathbb{R}^{(n-1)\times (n-1)}$ is a diagonal matrix with all the diagonal elements being the 
nonzero
eigenvalues of the matrix $\bm L_a$. 
With these block matrices,  it yields from (\ref{transformed}) that 
\begin{align}
\dot{\bm x}_2(t)=-\bm\Lambda_{n-1} \bm x_2(t)+\bm U_2^\top \bm B\bm w(t).
\end{align}
The output $\widehat{\bm y}(t)$ becomes 
\begin{align*}
\widehat{\bm y}(t)=\bm C^\top \widehat{\bm\varphi}&=\bm C^\top\bm U\bm x(t)
\\
&=
\begin{bmatrix}
\bm C^\top \bm u_1&\bm C^\top \bm U_2
\end{bmatrix}
\bm x(t)=
\bm C^\top \bm U_2 \bm x_2(t),
\end{align*}
where $\bm C^\top \bm u_1=\eta\bm C^\top \bm 1=\bm 0$ is used.  Hence,
 the output is independent of the component $x_1$. 
Because the  system matrix, which equals to $-\bm\Lambda_{n-1}$, is Hurwitz, there exists an invariant probability 
distribution for the state $\bm x_2(t)$, with the expectation $\bm m_{x_2}=0$ and the 
variance matrix $\bm Q_2=(q_{2,ij})\in\mathbb{R}^{(n-1)\times(n-1)}$ satisfying the Lyapunov equation
\begin{align}\label{Lyapunov}
\bm 0=-\bm \Lambda_{n-1}\bm Q_2-\bm Q_2\bm\Lambda_{n-1}+\bm U_2^\top \bm B\bm B^\top\bm U_2,
\end{align}
From the above equation, we further derive the analytic solution of $\bm Q_2$, 
\begin{equation}\label{Q_2solution}
    q_{2,ij}=(\lambda_{i+1}+\lambda_{j+1})^{-1}\bm u_{i+1}^{\top}\bm B\bm B^{\top}\bm u_{j+1}, i,j=1,2,\cdots,n-1.
\end{equation}
Because of the dependence on the state $\bm x_2$, there also exists an invariant probability distribution for the output $\widehat{\bm y}(t)$
with the expectation $\bm m_{\widehat{y}}=\bm 0$ and variance matrix 
\begin{equation}
    \label{variance matrix}
\bm Q_{\widehat{y}}=\bm C^{\top}\bm U_2\bm Q_2\bm U_2^{\top}\bm C.
\end{equation}
\par 
Next, we consider the first hitting time of the state hitting 
the boundary of the secure domain in the system (\ref{Stochastic Equation}). Clearly,  in a fixed interval of time, the higher the probability that the state stays in the secure domain (\ref{security condition}), the larger the mean first hitting time is. 
Here, instead of the probability density function of the non-linear stochastic process (\ref{Stochastic Equation}), we
 focus on the invariant probability distribution of a linear stochastic process, which is defined as 
 \begin{align}
 \widetilde{\bm y}&=\widehat{\bm y}(t)+\bm y^*,\label{stochasticoutput}
\end{align}
where $\widehat{\bm y}$
is the output of the system (\ref{linear System}). 
It is remarked that $\widetilde{\bm y}(t)$ approximates $\bm y(t)$ at the neighborhood of $\bm y^*$ due to the linearisation 
of the system (\ref{Stochastic Equation}) at the synchronous state $\bm\varphi^*$ with $\bm w(t)$ dealt as an input to the system.
Because $\bm y^*$ is a constant vector, the stochastic process $\widetilde{\bm y}$
is also Gaussian such that 
\begin{align}
&\widetilde{\bm y}(t)\in G\big(\bm m_{\widetilde{y}}(t),\bm Q_{\widetilde{y}}(t)\big),\label{tilde_y}\\
&\bm m_{\widetilde{y}}(t)=\bm m_{\widehat{y}}(t)+\bm y^*,~~\bm Q_{\widetilde{y}}(t)=\bm Q_{\widehat{y}}(t).
\end{align}
Thus, there exists an invariant probability for the Gaussian process $\widetilde{\bm y}(t)$ in (\ref{tilde_y}) with 
\begin{align*}
&\bm m_{\widetilde{y}}(t)=\bm y^*,~~\bm Q_{\widetilde{y}}(t)=\bm Q_{\widehat{y}}, \forall t\in \mathbb{T}. 
\end{align*} 
\par 
If $\widetilde{\bm y}(0)\in G(\bm y^*,\bm Q_{\widehat{y}})$, the process of $\widetilde{\bm y}(t)$ is 
a stationary process, in which $\widetilde{\bm y}(t)$ fluctuates around its expectation $\bm y^*$ with variance 
matrix $\bm Q_{\widehat{y}}$. If the $\widetilde{\bm y}(0)\notin G(\bm y^*,\bm Q_{\widehat{y}})$,
the distribution of $\widetilde{\bm y}(t)$ will converge to the invariant distribution $G(\bm y^*,\bm Q_{\widehat{y}})$. 
Note that with sufficient small disturbances in a short time period,
the process $\bm y(t)$ defined by (\ref{Stochastic Equation}) and (\ref{output-linear}) also fluctuates in the neighborhood of $\bm y^*$. 
Because $\widetilde{\bm y}(t)$ is an approximation of $\bm y(t)$, the variance matrix $\bm Q_{\widehat{y}}$
can be used to characterize the magnitude of the fluctuations of $\bm y(t)$. 
\par 
As shown in (\ref{variance matrix}), the variance
matrix of the phase difference is determined by the network 
topology and the spectrum of the Laplacian matrix.
Due to the dependence of the Laplacian matrix on the natural frequency 
and the coupling strength, the variance matrix also depends on these 
parameters. In addition, in contrast to the expectation $\bm y^*$, the variance matrix $\bm Q_{\widehat{y}}$ depends on the strength of the noise.  Here, the trace of the matrix $\bm Q_{\widehat{y}}$ is the $\mathcal{H}_2$ norm of the linear system (\ref{linear System}) where $\bm w(t)$ is seen as an input to the system.
This $\mathcal{H}_2$ norm is often used to analyse the fluctuations of the system subjected to disturbances\cite{H2norm,optimal_inertia_placement,RobustnessSynchrony}. See the Appendix \ref{InvariantDistribution} for details of the $\mathcal{H}_2$ norm. 
\par 

\par 
\begin{remark}
Instead of the system (\ref{linear System}), the following linear system
\begin{equation}\label{linearinputoutput}
\begin{aligned}
    \dot{\overline{\bm\varphi}}(t)&=\bm \omega-\bm L\overline{\bm\varphi}(t)+\bm B\bm w(t),\\
     \overline{\bm y}(t)&=\bm C^{\top}\overline{\bm\varphi}(t),
\end{aligned}
\end{equation}
may be studied to improve the synchronization stability of the system (\ref{deterministicmodel}) by increasing the probability that the state remains in the secure domain according to the invariant probability distribution. Here, $\bm \omega=\text{col}(\omega_i)\in\mathbb{R}^n$, $\bm L\in\mathbb{R}^{n\times n}$ is the Laplacian matrix of the weighted graph with weight $l_{ij}$ for the edge $(i,j)\in\mathcal{E}$, which is different from the matrix $\bm L_a$ defined in  (\ref{linear System}), $\bm B$ and $\bm C$ are the same as the ones defined in (\ref{linear System}). 
This system is derived from (\ref{Stochastic Equation}) by replacing the term $\sin(\varphi_i-\varphi_j)$ by $(\varphi_i-\varphi_j)$ directly. 
\par 
In fact,  the expectation of the phase difference satisfies 
\begin{align}\label{linearstochastic}
\overline{\bm{y}}^*=\bm C^\top \bm L^{\dagger}\bm\omega
\end{align}
where $\bm L^{\dagger}$ is the Moore-Penrose inverse of $\bm L$.  The matrix $\bm L$
has the following spectral decomposition 
\begin{align*}
\bm V^\top\bm L\bm V=\bm\Gamma,
\end{align*}
where $\bm \Gamma=\text{diag}(\gamma_i)\in\mathbb{R}^{n\times n}$ with $\gamma_i$ is the eigenvalue 
of the matrix $\bm L$ and the column vector $\bm v_i$ of $\bm V$ is the 
eigenvector of $\bm L$ corresponding to the eigenvalue $\sigma_i$. Since $\bm L\bm 1=\bm 0$, $\gamma_1=0$ is 
an eigenvalue with eigenvector $\bm v_1=\tau\bm 1$. Similar to the matrix $\bm U$ in (\ref{spectral}),
$\bm V$ is rewritten as $\bm V=[\bm v_1,\bm V_2]$ with $v_1=\tau \bm 1\in\mathbb{R}^n$ and 
$\bm V_2=[\bm v_2,\cdots,\bm v_n]\in\mathbb{R}^{n\times (n-1)}$. The variance matrix of the output satisfies
\begin{align}\label{linearsystemstochastic}
\overline{\bm Q}_y&=\bm C^\top \bm V_2\overline{\bm Q}_2\bm V_2^\top\bm C,
\end{align}
 where $\overline{\bm Q}_2=(\overline{q}_{2,ij})\in\mathbb{R}^{(n-1)\times(n-1)}$ such that 
\begin{equation}
    \overline{q}_{2,ij}=(\gamma_{i+1}+\gamma_{j+1})^{-1}\bm v_{i+1}^{\top}\bm B\bm B^{\top}\bm v_{j+1}, i,j=1,\cdots,n-1.
\end{equation}
The invariant probability distribution can also be obtained with the expectation in (\ref{linearstochastic}) and the variance in (\ref{linearsystemstochastic}).
\par 
However, in the invariant distribution, $\overline{\bm y}(t)$ fluctuates 
around $\overline{\bm y}^*$ calculated from (\ref{linearstochastic}), which is obviously different from $\bm y^*$ at the synchronous state 
(\ref{synchronous state}) and this difference increases as the 
synchronous state of the system (\ref{deterministicmodel}) moves to the boundary of the secure domain. In addition, because of the independence of $\bm V$ and $\bm\Gamma$ on
the synchronous state, the variance matrix $\overline{\bm Q}_{y}$ in (\ref{linearsystemstochastic}) is independent on the synchronous state.
This is different from the variance matrix $\bm Q_{\widehat{y}}$ in (\ref{variance matrix}), which depends
the eigenvalues of the matrix $\bm L_a$ that yields from the linearization at the synchronous state. 
Due to this independence, the nonlinearity of the system (\ref{deterministicmodel})
cannot be reflected by the probability of the state being absent from the secure domain 
at the invariant probability distribution of the process (\ref{linearinputoutput}). 
\end{remark}

\section{The optimization framework}\label{sec:optimization}

To increase the mean first time of the state hitting the boundary of the secure domain,
a way is to increase the probability of the state staying in the domain.
Since $\widetilde{\bm y}$ is an approximation of $\bm y(t)$ at the neighborhood of $\bm y^*$
and the distribution of $\widetilde{\bm y}(t)$ will converge to its invariant distribution, 
we focus on the probability of the process $\widetilde{\bm y}(t)$ staying 
in the secure domain in the invariant distribution.  However,
this probability can hardly be computed in practice due to an integral over a supercube of dimension $m$, which 
involves immense computational complexity.
Thus, we focus on the components of $\widetilde{\bm y}(t)$, which are
the stochastic process of the phase differences in the edges. 
In the invariant probability distribution, for edge $e_k$, the expectation and the variance of the phase difference
are denoted by $\mu_k$ and $\sigma_k^2$ respectively, 
which are computed as 
\begin{align}
\mu_k=y_k^*, ~\sigma_k^2=q_{kk}, ~\text{for}~ k=1,\cdots, m,\label{variance-diag}
\end{align}
where $y_k^*$ is the phase difference at the synchronous state that can be calculated from (\ref{expectation}), 
$q_{kk}$ is the $k$-th diagonal element of the matrix $\bm Q_{\widehat{y}}$ which is solved from (\ref{variance matrix}). 
The probability that the phase difference $\widetilde{y}_k(t)$ in edge $e_k$ belongs to the secure domain according to the invariant probability
distribution is
\begin{equation}\label{stable-stability}
    s_k(\mu_k,\sigma_k)=\int_{-\frac{\pi}{2}}^{\frac{\pi}{2}}
    \frac{1}{\sigma_k\sqrt{2\pi}}e^{\frac{-(x-\mu_k)^2}{2\sigma_k^2}}\text{d}x,
\end{equation}
Hence, the probability according to the invariant probability distribution that
the phase difference of the process $\widetilde{y}_k(t)$ is outside the secure domain, is equal to,
\begin{equation}
    \label{unstable probability}
    p_k(\mu_k,\sigma_k)=1-s_k(\mu_k,\sigma_k),
\end{equation}
for edge $e_k$ for $k=1,\cdots,m$.  Due to the approximation of the process (\ref{stochasticoutput}) to the output process
of system (\ref{Stochastic Equation}), this value measures
the risk of the phase difference in edge $e_k$ of the system (\ref{Stochastic Equation}) exceeding $\pi/2$. Thus, by this value, the vulnerable edges at which the system loses
the synchronization can be identified. Based on this value, we use the $\mathcal{L}_\infty$ 
norm of the vector $\bm P(\bm \mu,\bm \sigma)$ to measure the risk of the state hitting the boundary
of the secure domain, i.e.,
\begin{subequations}\label{norm}
\begin{align}
\bm P(\bm \mu,\bm \sigma)=\text{col}\big(p_k(\mu_k,\sigma_k)\big)\in\mathbb{R}^{m}, \\
||\bm P(\bm \mu,\bm \sigma)||_\infty=\max_{k=1,\cdots,m}\{p_k(\mu_k,\sigma_k)\},
\end{align}
\end{subequations}
where $\bm \mu=\text{col}(\mu_k)\in\mathbb{R}^m$ and $\bm \sigma=\text{col}(\sigma_k)\in\mathbb{R}^m$.  
Clearly, the risk of losing synchronization increases as the probability of the phase difference 
presenting outside of the secure domain. Thus, this norm also measures the risk of the system 
losing the synchronization. 
\par 
The following proposition describes the ranges of this norm and its relationship with the second smallest eigenvalue 
of the matrix $\bm L_a$, which is often used to study the linear stability of the complex system (\ref{Kuramoto Equation})\cite{pecora,Dorfler20141539}. 
\begin{proposition}\label{proposition}
Consider the invariant probability distribution of the processes $\widetilde{\bm y}(t)$ of the phase differences in the edges defined in (\ref{stochasticoutput}). It holds that 
\begin{enumerate}[(1)]
\item the norm $||\bm P(\bm \mu,\bm \sigma)||_\infty$ ranges over the interval $[0,1]$, 
\item if the second smallest eigenvalue of the matrix $\bm L_a$ decreases to zero, then the norm $||\bm P(\bm \mu,\bm \sigma)||_\infty$ defined in (\ref{norm}) increases to the value 1.
\end{enumerate}
\end{proposition}
\textbf{Proof:} (1) At a synchronous state, when the 
strength of the disturbances vary from zero to infinity, for $k=1,\cdots, m$, the variance $\sigma_k$
for edge $e_k$ varies from zero to infinity. Following from (\ref{stable-stability}-\ref{unstable probability}),
the range of $p_k(\mu_k,\sigma_k)$ is $[0,1]$, thus this norm also lies in $[0,1]$. 
\par 
(2) For $\bm A,\bm B\in\mathbb{R}^{n \times n}$, we say that $\bm A\preceq \bm B$ if the matrix $\bm A-\bm B$ is semi-negative definite. 
Define $\underline{b}=\min\{b_i,i=1,\cdots,n\}$. Then,
\begin{align*}
\bm B\bm B^\top\succeq \underline{b}^2\bm I_n,
\end{align*}
where $\bm I_n\in\mathbb{R}^{n\times n}$ is an identity matrix. From (\ref{Lyapunov}) and (\ref{variance matrix}),
we derive
\begin{align*}
\bm Q_2\succeq\frac{1}{2}\underline{b}^2\bm\Lambda_{n-1}^{-1},~~\bm Q_{\widehat{y}}\succeq \frac{1}{2}\underline{b}^2 \bm C^\top\bm U_2\bm\Lambda_{n-1}^{-1}\bm U_2^\top\bm C.
\end{align*}
To prove this proposition, we only need to prove that, as the second smallest eigenvalue decreases
to zero, there is at least one diagonal element of the matrix $\bm S=\bm C^\top\bm U_2\bm\Lambda_{n-1}^{-1}\bm U_2^\top\bm C$ that increases to infinity. The incidence matrix of the 
graph is written into 
$\bm C=\begin{bmatrix}
\bm c_1&\bm c_2&\cdots&\bm c_m
\end{bmatrix}
$ where the vector $\bm c_k $ describes the indices of the two nodes that are connected by edge $e_k$. Without losing
generality, assume the direction of edge $e_k$  is from node $i$ to $j$.
Then, the $i$-th and $j$-th element of the vector $\bm C_k$, $c_{ik}=1$ and $c_{jk}=-1$, respectively and the other elements 
all equal to zero. From the definition of the matrix $\bm S$, we obtain the diagonal element of $\bm S$, 
\begin{align*}
s_{kk}=\sum_{q=1}^{m-1}\lambda_{q+1}^{-1}(u_{i,q+1}-u_{j,q+1})^2,~~k=1,2\cdots m,
\end{align*}
where $u_{i,q+1}$ and $u_{j,q+1}$ are the $i$-th and $j$-th element of the vector $\bm u_{q+1}$ and $\bm u_{q+1}$ is the 
$(q+1)$-th column of the matrix $\bm U$ defined in (\ref{spectral}). 
Because $\bm u_2$ is the second column of the orthogonal matrix $\bm U$, which is the 
eigenvector of $\bm L_a$ corresponding to the second smallest eigenvalue $\lambda_2$, there exist $i,j$ with $i\neq j$ 
such that $u_{i,2}\neq u_{j,2}$, thus $s_{kk}$ increases to infinity as the second smallest eigenvalue $\lambda_2$ decreases to zero.
\hfill $\square$

As the natural frequencies increasing or the coupling strength of the edges decreasing, the synchronous state will move
towards the boundary of the secure domain. In this case, 
the basin of attraction of the synchronous state gradually disappears and 
the second smallest eigenvalue of $\bm L_a$ decreases to zero. 
Because 
the norm $||\bm P(\bm \mu,\bm \sigma)||_\infty$ increases to its upper bound as the second smallest eigenvalue of $\bm L_a$ decreases 
to zero when the synchronous state disappears, it fully indicates the response of the linear stability and nonlinear stability to these system parameters. 
Besides this property, the value $\bm P(\bm \mu,\bm \sigma)$ also
depends on the strength of the noise due to the dependence of $\bm \sigma$ on the strength of the noise. This is different from the spectrum of the system matrix and 
the size of the basin of the attraction, which are independent of the strength of the noise. 
In addition, 
with the elements in the vector $\bm P(\mu,\sigma)$, the response 
of the vulnerability of each edge to the changes of the system parameters can be captured. Thus,
this metric is more practical and comprehensive for the analysis of the synchronization stability. 
\par

With the metric $||\bm P(\bm \mu,\bm \sigma)||_\infty$, we propose an optimization framework to
increase the mean first hitting time thus enhance the synchronization stability. 
In this optimization framework, the objective is minimizing the risk of the state hitting the boundary of the secure domain and the decision variables include the coupling strength and the natural frequency.  
\par 
We first study the effects of the coupling strength given the natural frequency and the network topology.  It is well known that the synchronization stability increases as the coupling strength of the edges increase.  Thus, we consider the networks with a constant total amount of coupling strengths. Consider the system (\ref{Stochastic Equation}), 
the optimization problem for the assignment of the coupling strength is 
\begin{subequations}\label{optimizaiton}
\begin{align}
&\min_{l_{ij}\in\mathbb{R},(i,j)\in\mathcal{E}}||\bm P(\bm\mu,\bm\sigma)||_{\infty},\\
\nonumber
\text{s.t.} ~~&\text{(\ref{synchronous state}),(\ref{expectation}),(\ref{spectral}),(\ref{Q_2solution}),(\ref{variance matrix}),
(\ref{variance-diag}),} \\
&0=\sum_{(i,j)\in\mathcal{E}} {l_{ij}-W},\label{totalcouplingstrength}\\
\label{opitmizaion-b}
&\underline{l}_{ij}<l_{ij} <\overline{l}_{ij}~~\text{for}~~(i,j)\in\mathcal{E}.
\end{align}
\end{subequations}
where $W\in\mathbb{R}$ is 
the total amount of the coupling strength, $\underline{l}_{ij}>0$ and 
$\overline{l}_{ij}>0$ are respectively the lower and upper bounds of the coupling strength of the edge. 
In this optimization problem, the coupling strength does not only impact the synchronous state but also the variance of the phase differences, thus affects the synchronization stability in a non-linear way.  
\par
We next consider the assignment of the natural frequency given 
the coupling strength and the network topology.   Consider 
the system (\ref{Stochastic Equation}), the optimization problem for 
the design of the natural frequency is 
\begin{subequations}\label{optimizingnaturalfrequency}
\begin{align}
&\min_{\omega_i\in\mathbb{R},i\in\mathcal{V}}||\bm P(\bm\mu,\bm\sigma)||_{\infty},\\
\nonumber
\text{s.t.} ~~&\text{(\ref{synchronous state}),(\ref{expectation}),(\ref{spectral}),(\ref{Q_2solution}),(\ref{variance matrix}),
(\ref{variance-diag}),} \\
&0=\sum_{i=1}^n\omega_i,\label{constraintontotalfrequency0}\\
&\underline{\omega}_i<\omega_i<\overline{\omega}_i, i=1,\cdots,n.  \label{constraintontotalfrequency}
\end{align}
\end{subequations}
where $\underline{\omega}_i$ and $\overline{\omega}_i$ are 
the lower and upper bound of  $\omega_i$ respectively. 
\par 
In order to use the proposed metric $\bm P(\bm \mu,\bm \sigma)$ to analyze the synchronization stability
of complex systems in practice, one has to solve the nonlinear equations (\ref{synchronous state}) for the expectation
$\bm \mu$ and perform the matrix spectral decomposition (\ref{spectral}) for the variance matrix $\bm Q_{\widehat{y}}$, for which the Newton iterative method and the QR method can be used respectively. In particular, if the QR method
is used for the matrix decomposition, the estimated computing complexity is  $O(n^3)$. 
To solve the corresponding optimization problems, iterative methods can be used, where the solution 
of (\ref{synchronous state}) and  the matrix spectral decomposition (\ref{spectral}) are needed in each iteration.  
Thus,  besides efficient algorithms for solving the non-linear equation (\ref{synchronous state}) and for the matrix spectral decomposition (\ref{spectral}),  an iterative method for 
the optimization problems with a fast convergence rate is important for increasing the synchronization stability of large-scale systems using the proposed optimization framework.  
\par 

\section{Case study}\label{sec:casestudy}
We evaluate the performance of the 
optimization framework for increasing the synchronization stability.
Monte-Carlo method based numerical simulations are carried out to compute the mean first hitting
time of the nonlinear stochastic system (\ref{Stochastic Equation}) and to identify the vulnerable edges in the network. By these simulations,
we verify the effectiveness of the metric $p_k$ in (\ref{unstable probability}) on finding the vulnerable edges 
and of the optimization framework on increasing the first mean hitting time. 
\begin{figure}[ht]
    \centering
	\includegraphics[width=0.22\textwidth]{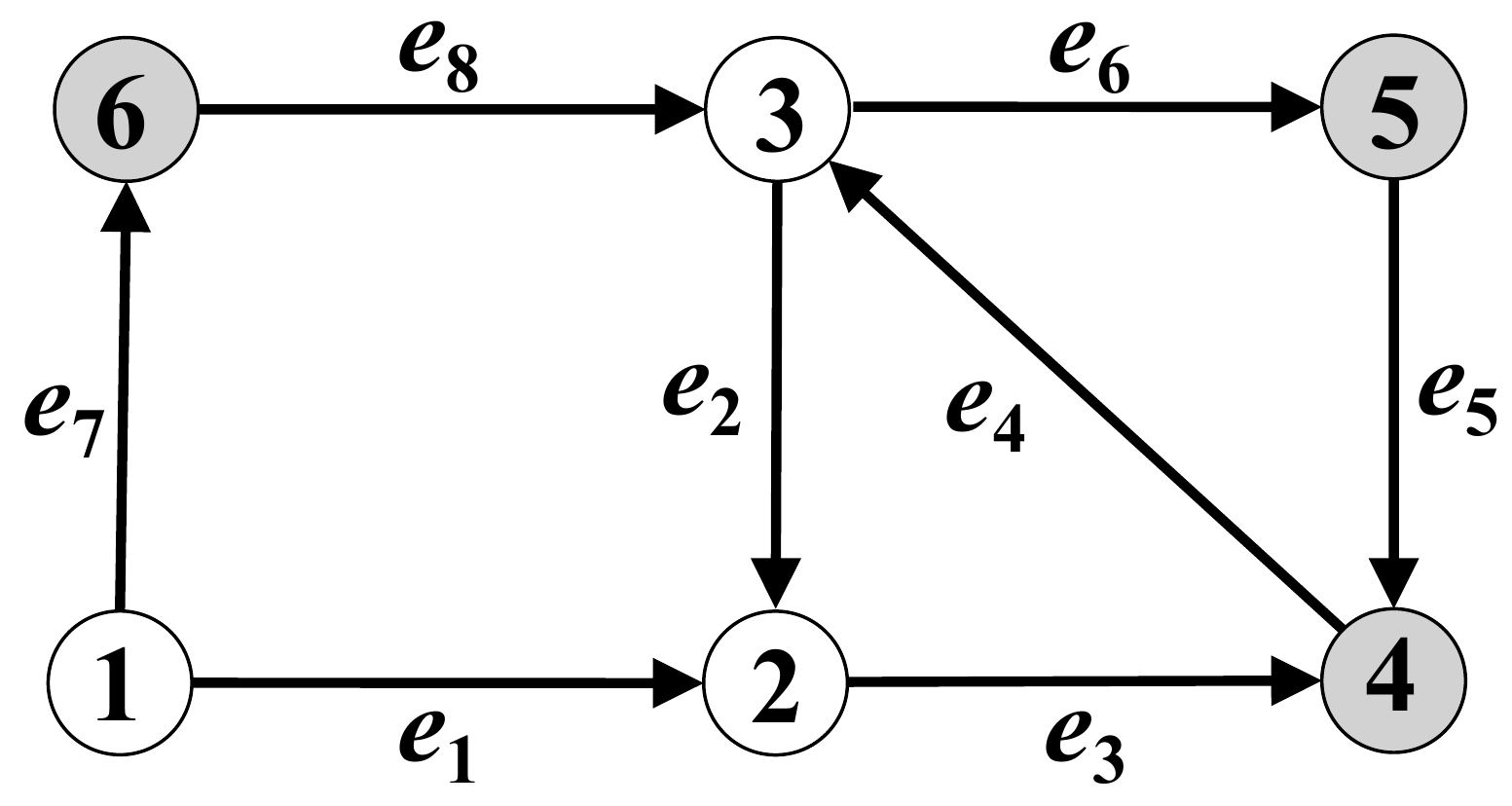}
    \caption{A network with 6 nodes and 8 edges. }\label{figure1}
\end{figure}
\par 
 In the simulations,  we use the Euler-Maruyama method to discretize the system (\ref{Stochastic Equation}) with the simulation time $T$, the time step size $\text{d}t$ and 
the initial condition $\bm \varphi(0)=\bm\varphi^*$. If there is an edge in which the absolute value of the phase difference exceeds $\pi/2$, the simulation is stopped. Then, the stopping time and the index of this edge are recorded. 
 The mean first hitting time $\overline{t}_e$ is obtained as the mean of the stopping time in these simulations. In these simulations, only those simulations are counted which lead to a stopped process within the simulation horizon $T$. The total number of the counted simulations is $N$ which almost equals to the total number of 
simulations. In addition, the number $g_k$ that the absolute value of the phase difference exceeding $\pi/2$ among the simulations is counted for the edge $e_k$. 
For each line, we calculate the following ratio 
\begin{align}\label{ratioRk}
r_k=g_k/N,
\end{align}
which satisfies $\sum_{k=1}^m r_k=1$. This ratio approximates the probability that the absolute value of the phase difference exceeds $\pi/2$ at line $e_k$ conditioned on that the state exits the secure domain.   
 Clearly, the larger the ratio for an edge, the easier the boundary of the secure domain is hit by the phase difference at this edge.
The risk of the phase difference exceeding $\pi/2$ at each edge is calculated from (\ref{unstable probability}). 
To compare with the ratio $r_k$, we calculate the value 
\begin{align}\label{vulnerbleLines}
\ds \widetilde{p}_k=\frac{p_k}{\sum_j^m{p_j}},~ \text{for}~k=1,\cdots,m,
\end{align}
which is the probability of the absolute value of the phase difference exceeding $\pi/2$ in edge $e_k$ conditioned on the state being absent from the secure domain in the invariant probability distribution of the linear stochastic process (\ref{stochasticoutput}). 
\par 
Regarding the effectiveness of the optimization framework in the 
enhancement of the synchronization stability, we compare the solutions of the following 5 optimization frameworks,
\begin{enumerate}[(1)]
\item Maximizing the order parameter $r$ at the synchronous state of the system (\ref{Kuramoto Equation}) \cite{Skardal2014}, 
see the optimization problems
(\ref{orderoptimization1}, \ref{orderoptimization2}) in Appendix \ref{optimization-Case-study};
\item Minimizing the $\mathcal{L}_\infty$ norm of the phase differences at the 
synchronous state, which aims to
increase the phase cohesiveness of the system (\ref{Kuramoto Equation}) \cite{FAZLYAB2017181},
see the optimization problems
(\ref{cohesivenessoptimization1},\ref{cohesivenessoptimization2}) in Appendix \ref{optimization-Case-study};
\item Minimizing the $\mathcal{L}_\infty$ norm of the variance of the phase differences in the invariant probability distribution of the process (\ref{stochasticoutput}), which aims to 
decrease the fluctuations in the phase differences of the system (\ref{Kuramoto Equation}) with disturbances. 
The corresponding optimization problems can be obtained by replacing the objective functions in (\ref{optimizaiton}) and (\ref{optimizingnaturalfrequency}) by $||\bm \sigma||_\infty$; 
\item Minimizing the $\mathcal{H}_2$ norm of the system (\ref{linear System}), which aims 
to decrease the fluctuations in the phase differences of the system (\ref{Kuramoto Equation}) with disturbances,
see the optimization problems
(\ref{H2normoptimization1}, \ref{H2normoptimization2}) in Appendix \ref{optimization-Case-study};
\item Minimizing the risk of the state hitting the boundary of the secure domain measured by $||\bm P||_\infty$, see the optimization problems (\ref{optimizaiton},\ref{optimizingnaturalfrequency}).
\end{enumerate}
In the first two optimization frameworks, the focuses of the objectives are on the synchronous state of the deterministic system (\ref{Kuramoto Equation}) where the impacts of the disturbances are not considered. However,
in the latter three optimization frameworks, the disturbances are involved in while the
synchronous state is not fully considered.  Note that by the metric of the phase cohesiveness, the vulnerable edges 
may be identified as the ones in which the phase differences are large, while by the metric of the variance of 
the phase difference, the vulnerable edges may also be identified as the ones 
in which the variances are large \cite{KaihuaXi2022}.  The optimization problems are solved by Matlab.
\par 

\begin{figure*}[ht]
  \centering
  \begin{subfigure}[b]{0.33\textwidth}
    \includegraphics[scale=0.33]{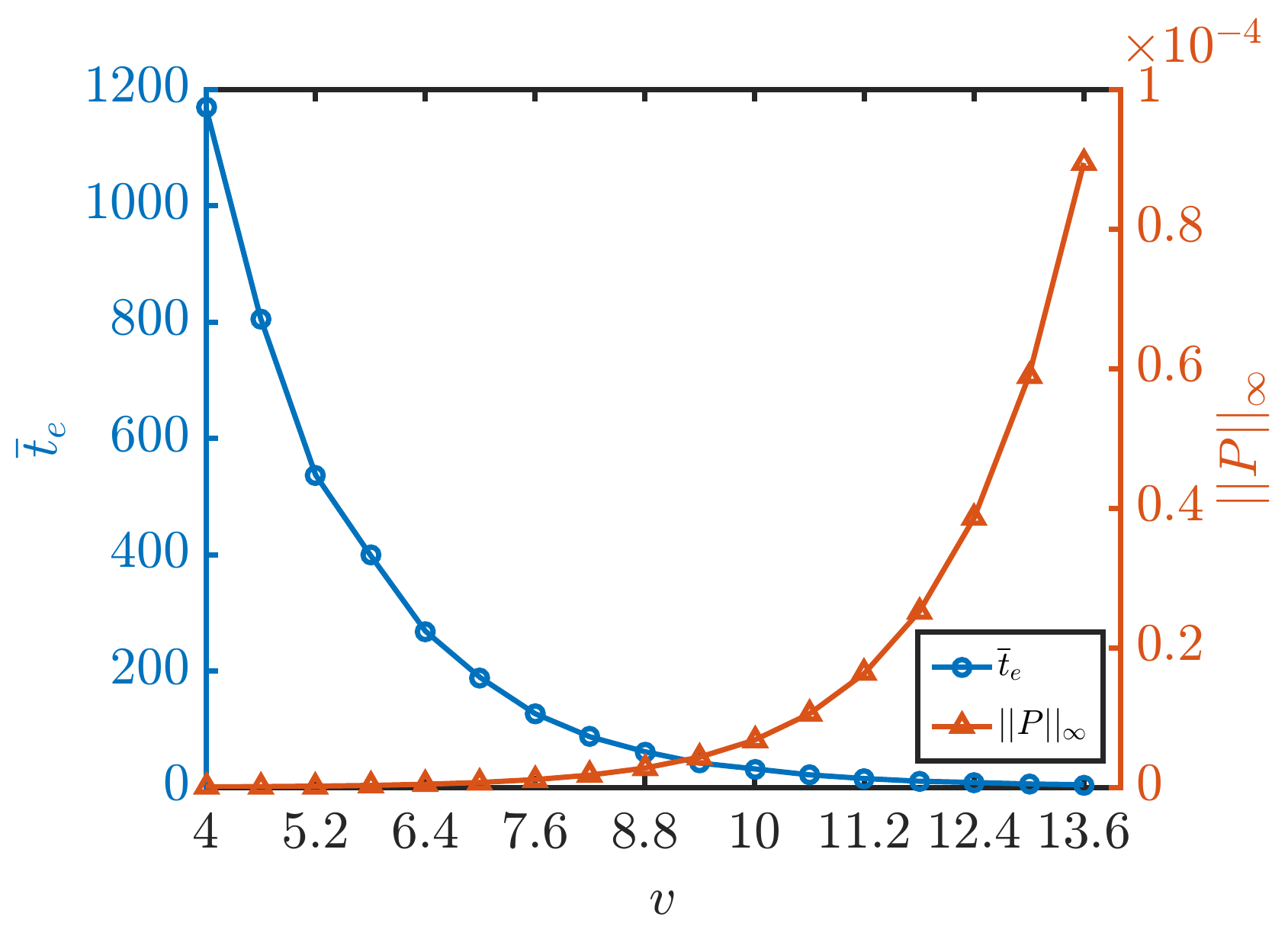}~~
    \subcaption*{(a)}
\end{subfigure}%
\begin{subfigure}[b]{0.33\textwidth}
  \includegraphics[scale=0.33]{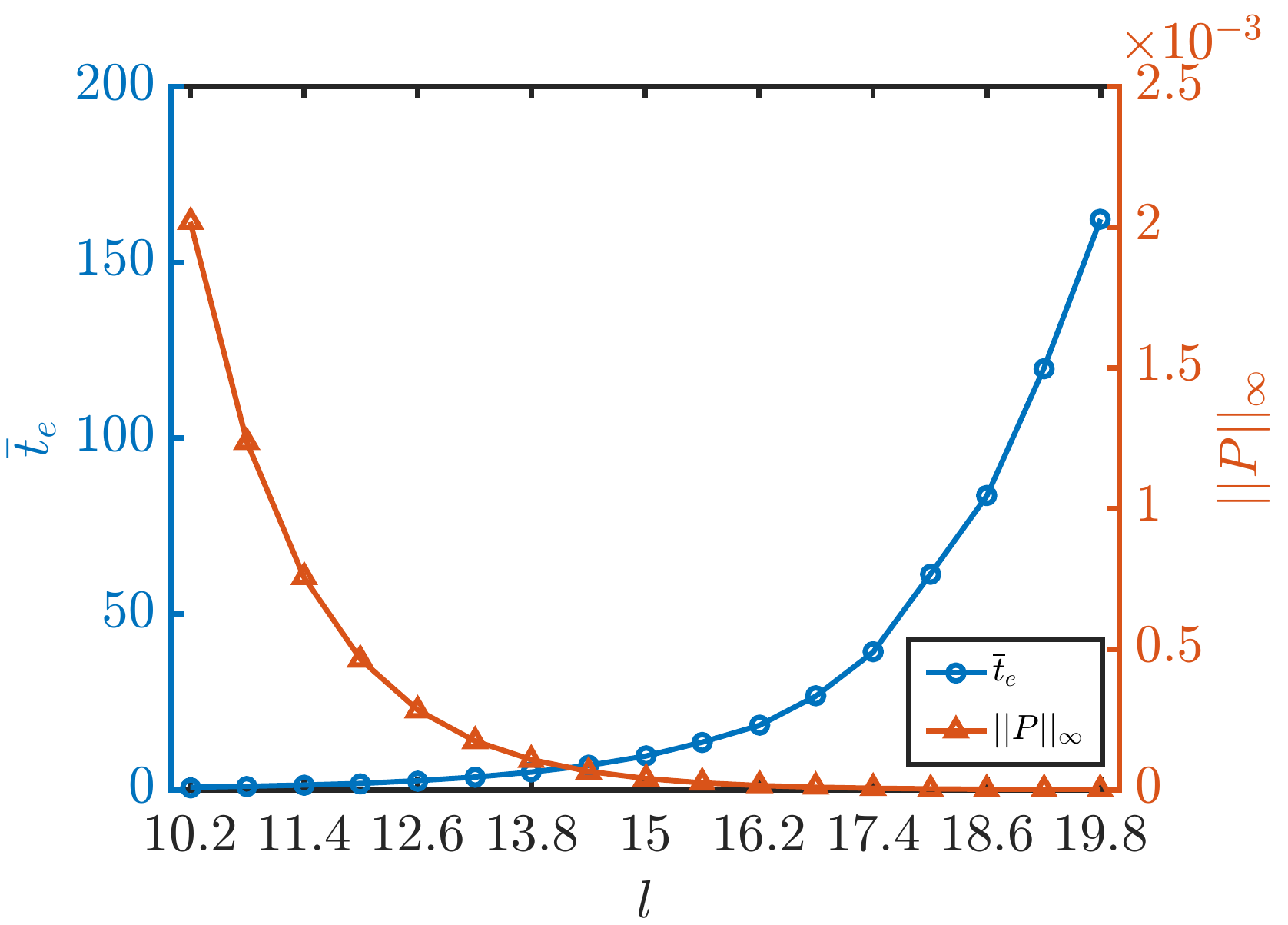}~~
  \subcaption*{(b)}
\end{subfigure}%
\begin{subfigure}[b]{0.33\textwidth}
  \includegraphics[scale=0.33]{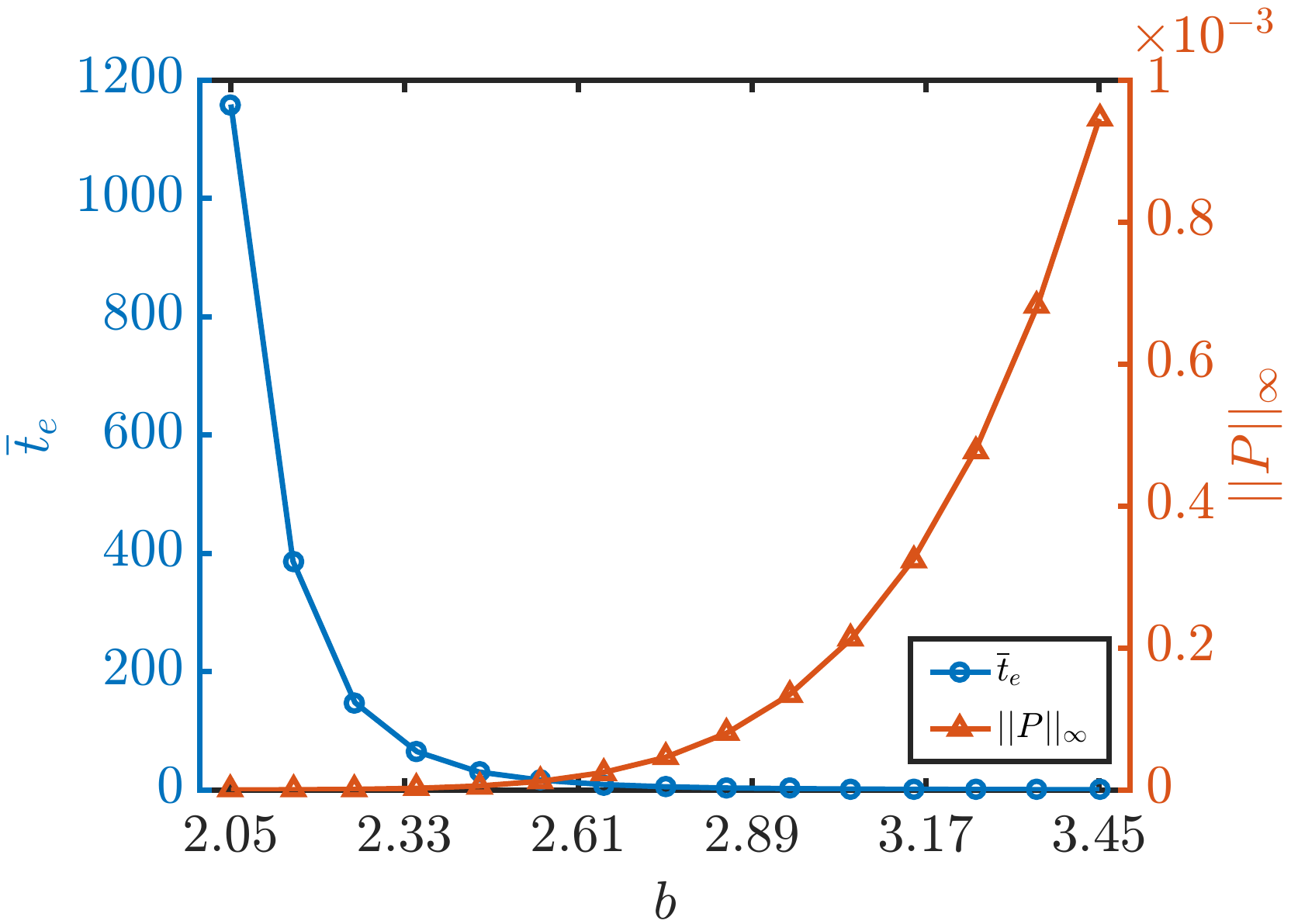}
  \caption*{(c)}
\end{subfigure}
\caption{The dependence of the mean first time $\overline{t}_e$ and the defined metric $\|P\|_\infty$ on the system parameters. (a) $\omega_i=v$ for $i=1,2,3$ and $\omega_i=-v$ for $i=4,5,6$ where $v$ is a positive constant, $l_{ij}=22$ for all the edges and 
$b_i=2.1$ for all the nodes. (b) $\omega=5$ for $i=1,2,3$ and $\omega=-5$ for $i=4,5,6$ and $l_{ij}=l$ for all 
the edges where $l$ is a positive constant, $b_i=2.1$ for all the nodes. (c) $\omega=5$ for $i=1,2,3$ and $\omega=-5$ for $i=4,5,6$ and $l_{ij}=22$ for all 
the edges where $b$ is a positive constant, $b_i=b$ for all the nodes.}
\label{objective-escapetime}
\end{figure*}
We evaluate the performance of the  proposed optimization framework for increasing 
the synchronization stability in the two networks shown in Fig. \ref{figure1} and Fig. \ref{figure-net2} respectively.
By the network in Fig. \ref{figure1}, we show the relationship between the metric $\|P\|_\infty$ and the mean 
first hitting time $\overline{t}_e$ and the performance of the vector $\bm P(\bm \mu,\bm\sigma)$ on identifying the vulnerable lines. In addition, by presenting the solutions of the corresponding optimization problems for the design 
of the coupling strength and the natural frequency, we show the performance of the proposed optimization framework on 
increasing the mean first hitting time. By the network in Fig. \ref{figure-net2}, we confirm the performance of the proposed optimization framework for relatively large scale systems. 

\par 

\par 
\begin{example}\label{example1}
Consider the network in Fig.\ref{figure1} with 6 nodes and 8 edges.  The natural frequencies at the grey nodes are negative while those at the other nodes are positive. The directions of the edges are specified arbitrary, which do not affect the analysis. We set $T=10^5$, $\text{d}t=10^{-3}$, $N=10^5$, $b_i=1.05$ for all the nodes. We formulate an Initial Model, in which we set $\omega_i=5$ for $i=1,2,3$, and $\omega_i=-5$ for $i=4,5,6$ and $l_{ij}=8$ for all the edges. 
In the optimization problems for the design of the coupling strength, we set $\omega_i=5$ for $i=1,2,3$, and $\omega_i=-5$ for $i=4,5,6$, the total coupling strength $W=64$ and $\underline{l}_{ij}=1$, 
$\overline{l}_{ij}=12$ for all the edges. 
In the optimization problems for the design of the natural frequency, we set 
$\underline{\omega}_i=-5$ and $\overline{\omega}_i=-5$ for nodes $4,5,6$ and 
$\underline{\omega}_i=0$ and $\overline{\omega}_i=15$ for nodes $1,2,3$, and $l_{ij}=8$ for all the edges.
\end{example}
\par 
We first focus on the relationship between the mean first hitting time and 
the risk of the state hitting the boundary of the secure domain measured by $||\bm P||_\infty$.   
Shown in Fig. \ref{objective-escapetime} are the dependence of the mean first hitting time and $||\bm P||_\infty$ on the natural frequency, the coupling strength and the disturbances. 
The configuration of the parameters are described in the caption Fig. \ref{objective-escapetime}. 
It is demonstrated that as the risk of the state hitting the boundary of the secure domain increases, the mean first hitting time
decreases. This indicates that the synchronization stability decreases.  
It can be imagined that as the risk of the state hitting the boundary of the secure domain increases to one, 
the mean first hitting time will decrease to zero. 
\begin{figure}[ht]
  \centering
\includegraphics[width=0.32\textwidth,height=0.21\textwidth]{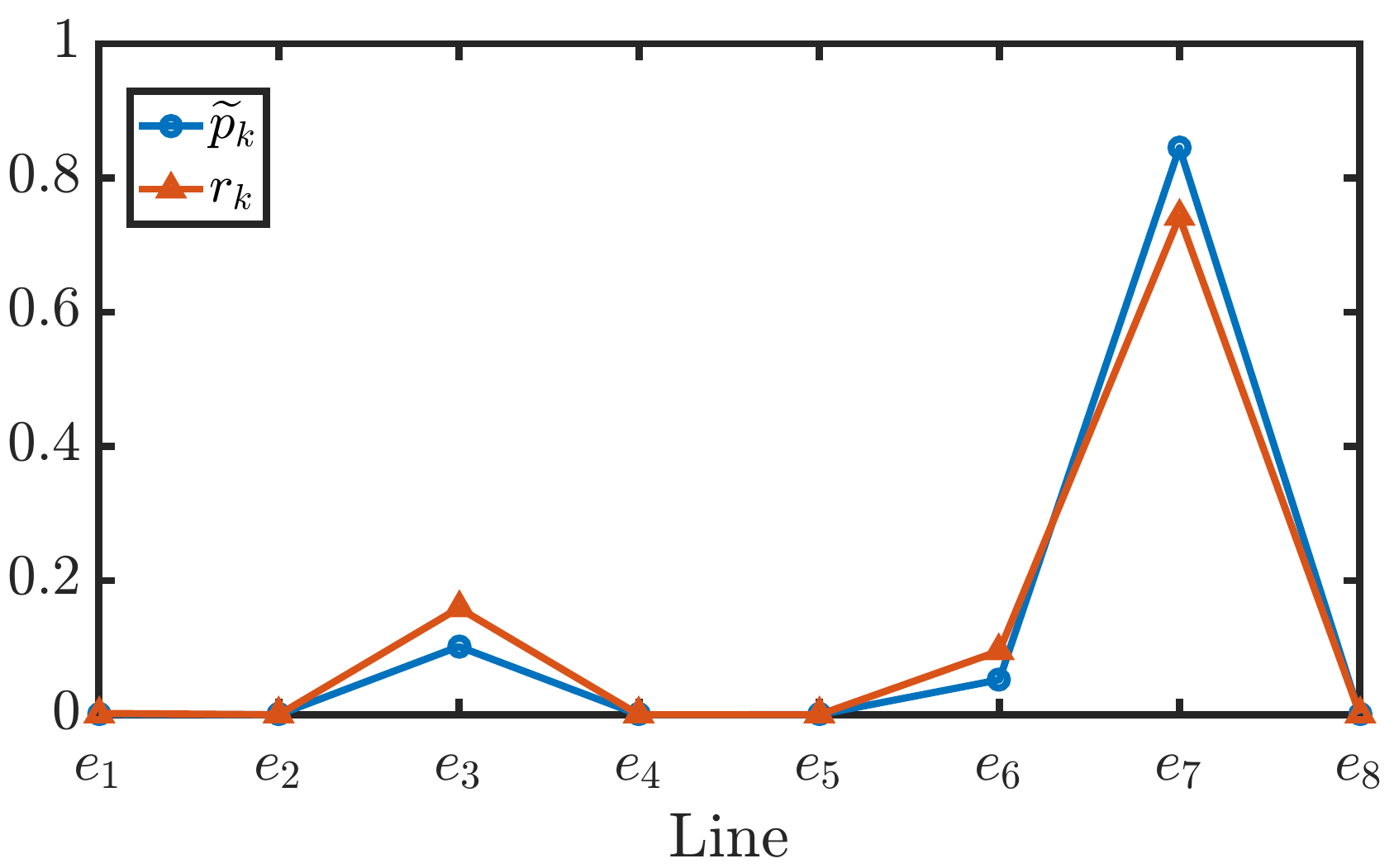}
  \caption{The value $\widetilde{p}_k$ and the ratio $r_k$ at the edges. We set $\omega_i=5$ for $i=1,2,3$, $\omega_i=-5$ for $i=4,5,6$, $l_{ij}=20$ for all the edges and $b_i=2.1$ for all the nodes.  }
  \label{unstable-probability}
\end{figure}
\par 
\begin{figure*}[ht]
  \centering
  \begin{subfigure}[b]{0.32\textwidth}
    \includegraphics[scale=0.36]{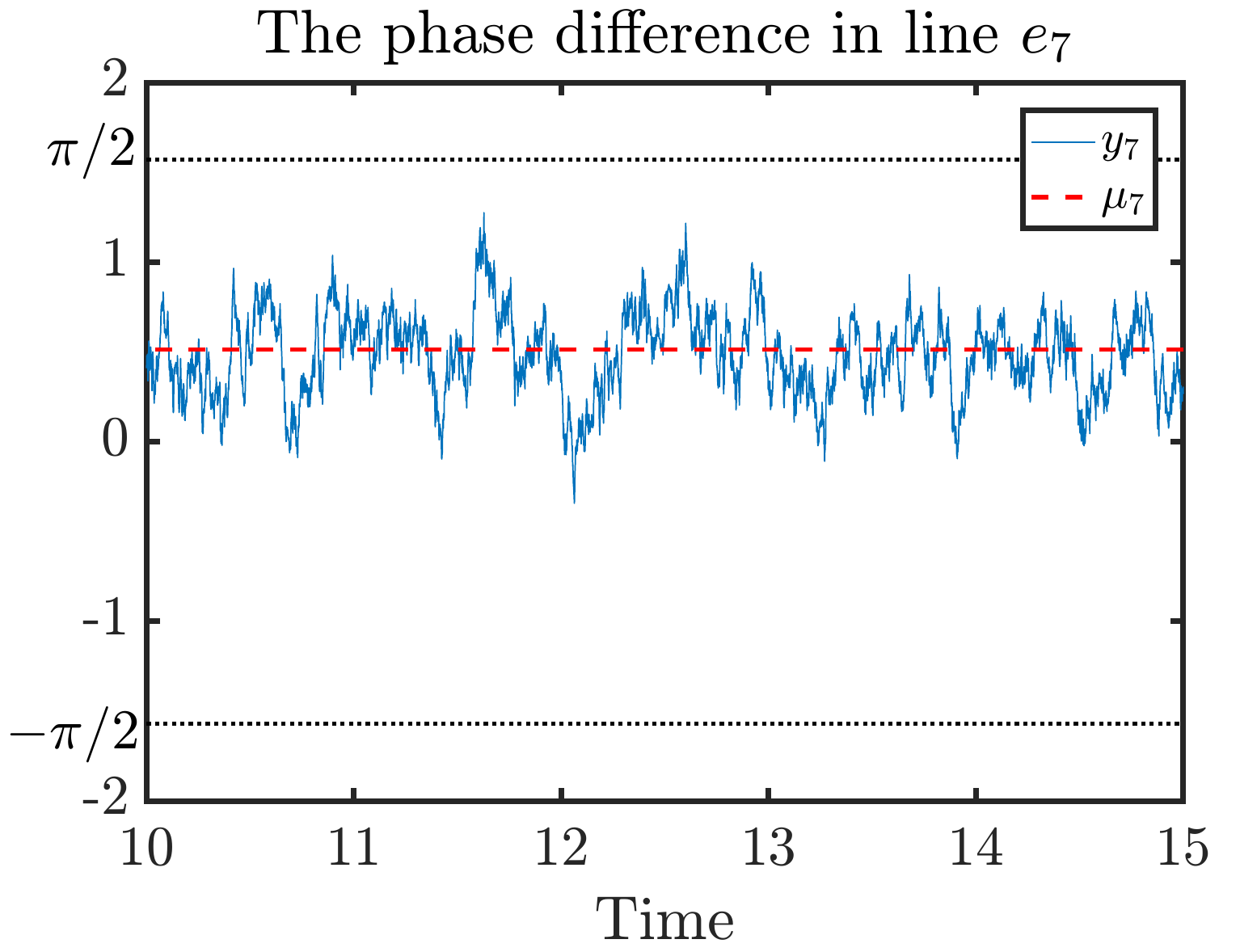}
    \caption*{(a)}
  \end{subfigure}
  \begin{subfigure}[b]{0.32\textwidth}
    \includegraphics[scale=0.36]{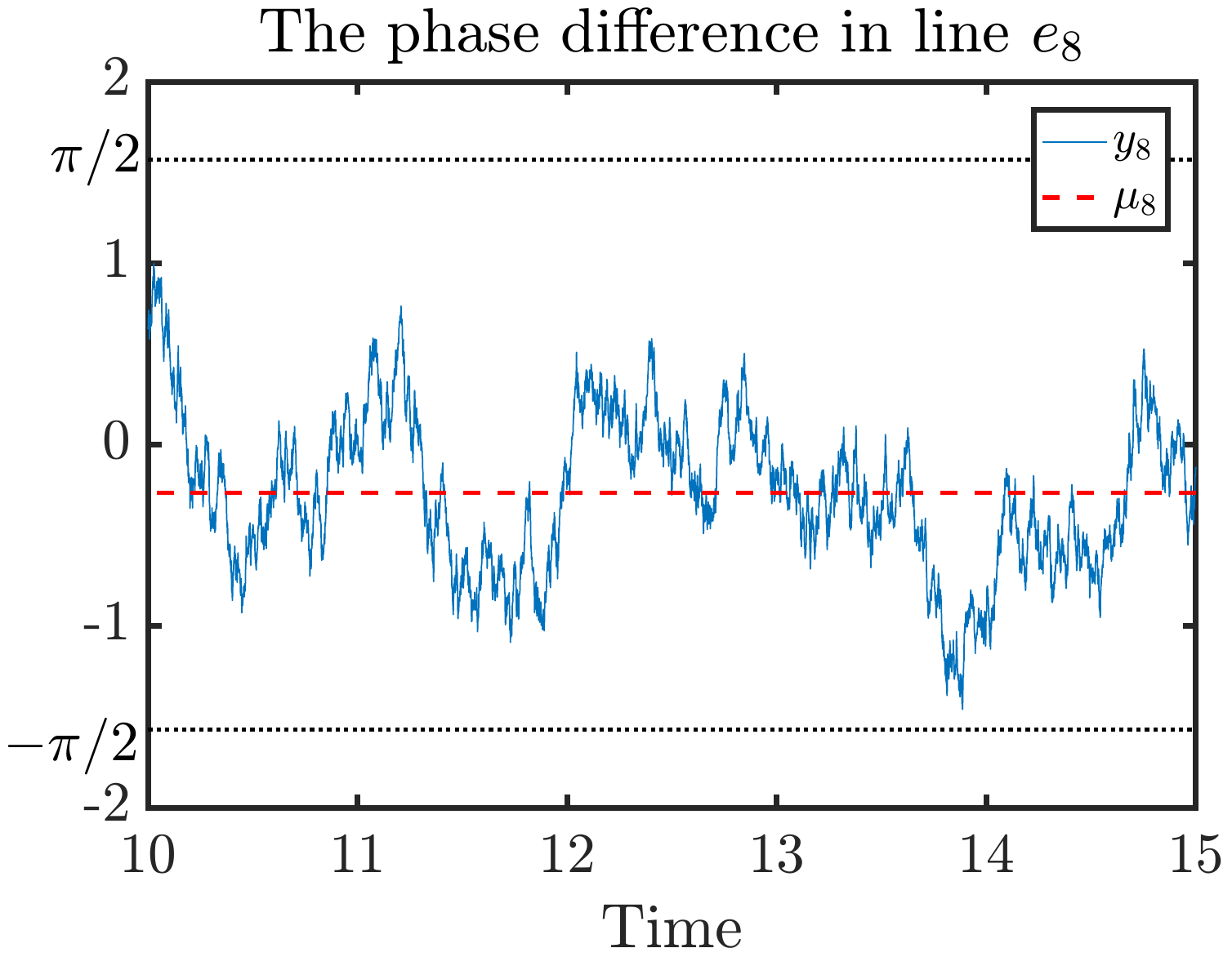}
    \subcaption*{(b)}
\end{subfigure}%
\begin{subfigure}[b]{0.32\textwidth}
  \includegraphics[scale=0.36]{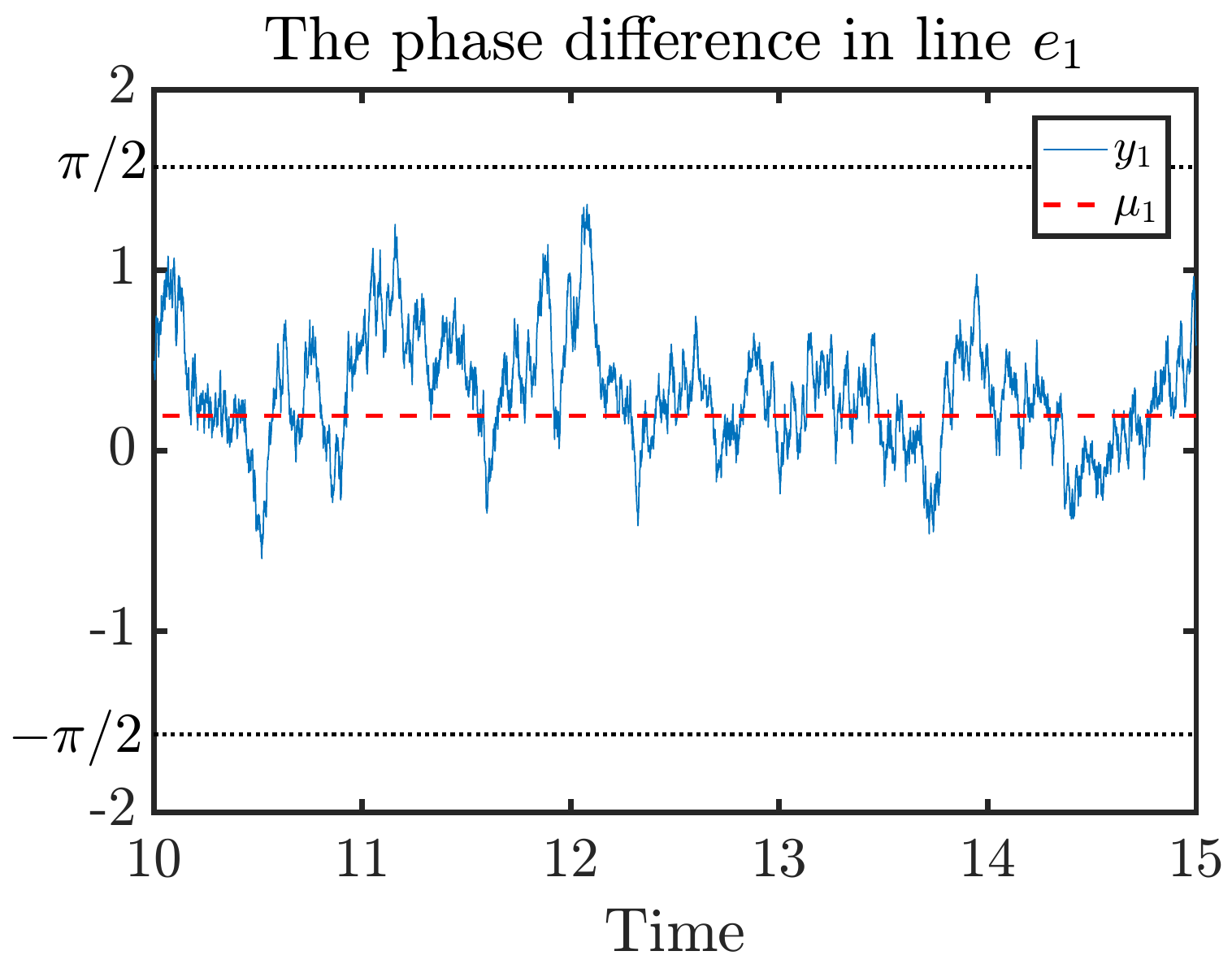}
  \subcaption*{(c)}
\end{subfigure}%
\par
\begin{subfigure}[b]{0.32\textwidth}
  \includegraphics[scale=0.36]{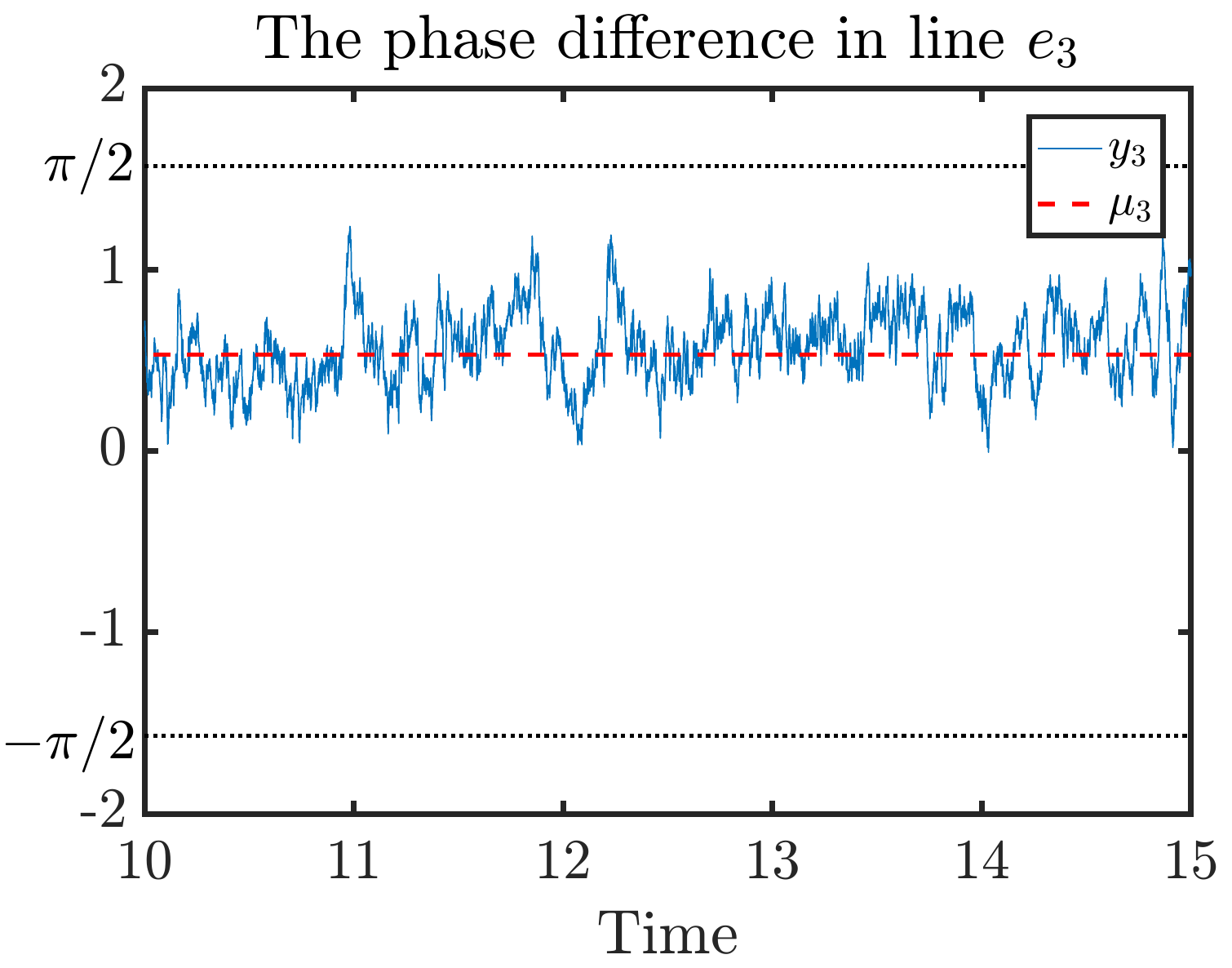}
  \caption*{(d)}
\end{subfigure}
\begin{subfigure}[b]{0.32\textwidth}
  \includegraphics[scale=0.36]{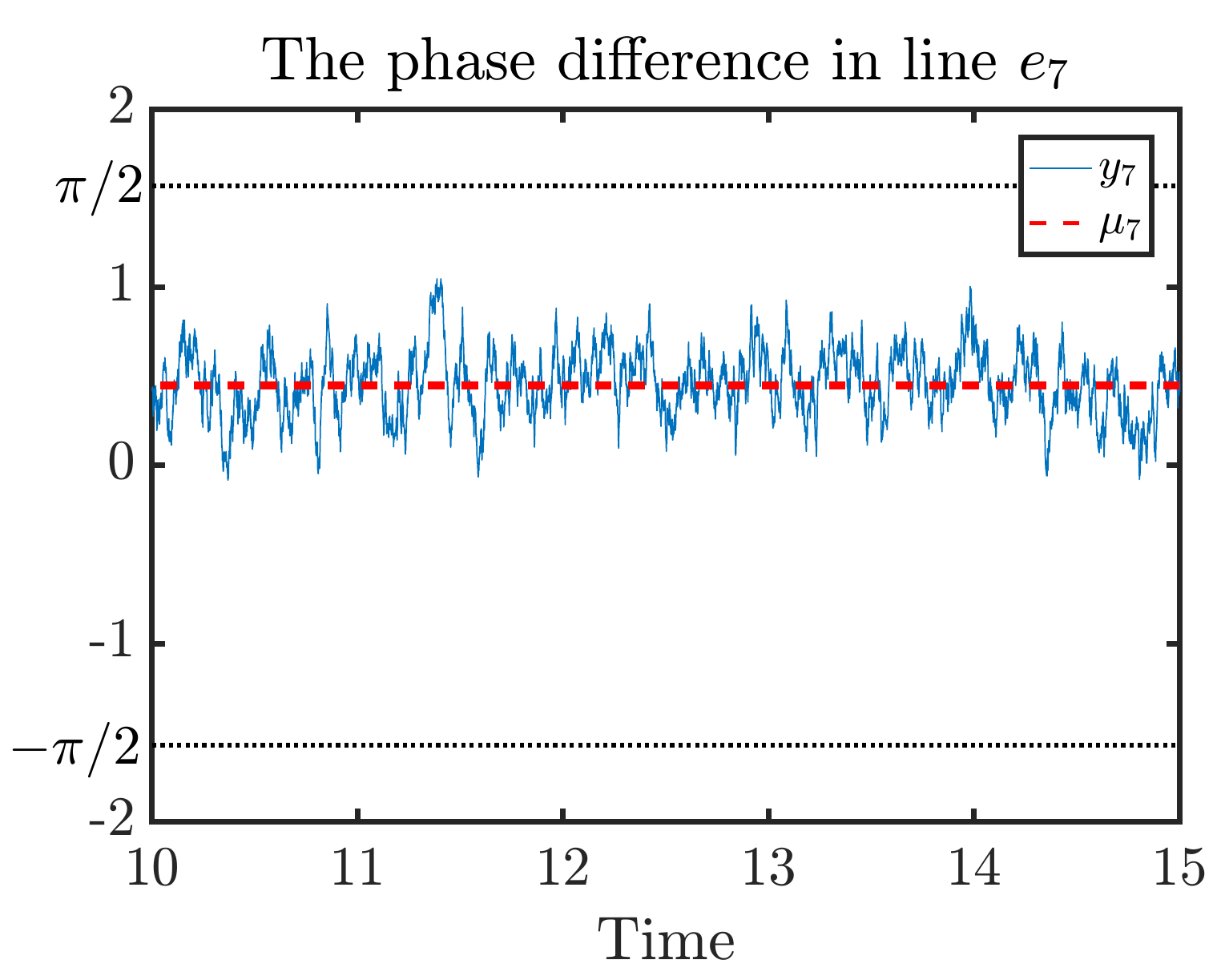}
  \subcaption*{(e)}
\end{subfigure}%
\begin{subfigure}[b]{0.32\textwidth}
  \includegraphics[scale=0.36]{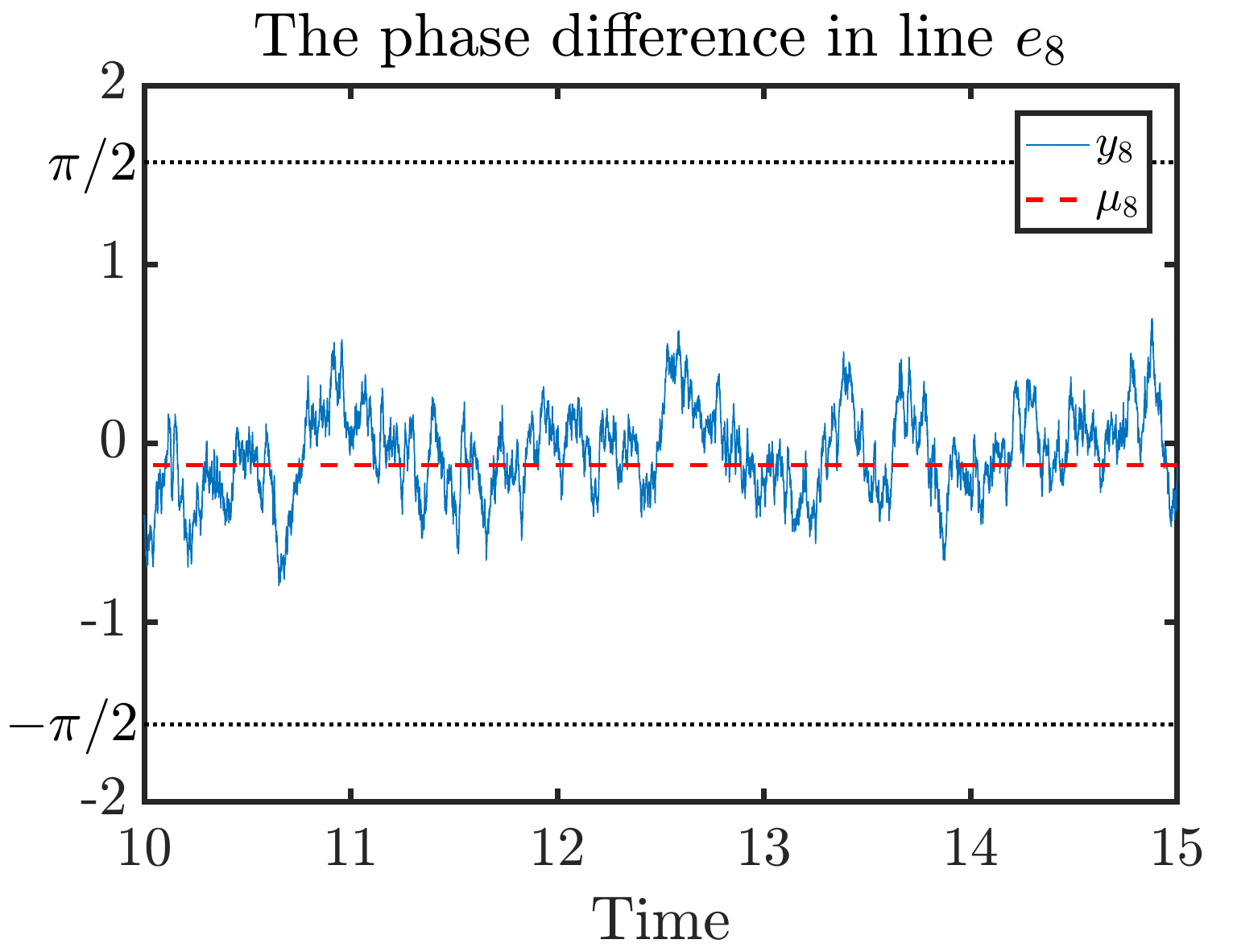}
  \caption*{(f)}
\end{subfigure}
\caption{The phase differences in the most vulnerable edges after designing the coupling strength with the 5 different objectives the network in Fig. \ref{figure1}. (a) Initial model, (b) Max. $\gamma$. (c) Min. $||\bm \mu||_\infty$. (d) Min. $||\bm \sigma||_\infty$. (e) Min. $\mathcal{H}_2$. (f) Min. $||\bm P||_\infty$.}
\label{objective1-result}
\end{figure*}

Next, we consider the identification of the vulnerable edges in the system (\ref{Stochastic Equation}) by 
the metric defined in (\ref{unstable probability}) in the network. 
The values $r_k$ and $\widetilde{p}_k$ for each edge are shown in Fig. \ref{unstable-probability}.  
It is demonstrated that $\widetilde{p}_k$ estimates $r_k$ well for all the edges and $e_7$ is the most vulnerable edge. Thus, the vulnerability of the edge can be measured by the metric $p_k$.
\par 

\par 
Let us investigate the optimal distribution of the coupling strength. 
Table \ref{design-coupling-strength} shows the optimal solution for the design of the coupling strength by the 
optimization problems with the 5 objectives.   
It can be seen that the mean first hitting time increases
from 118.46s to 363.396s, 773.220s and 3951.733s by minimizing the largest variance of the phase differences measured by $||\bm \sigma||_\infty$,
the $\mathcal{H}_2$ norm and the risk of the state hitting the secure domain measured by $||\bm P||_{\infty}$, respectively. It demonstrates that by suppressing the variance of the phase 
differences, i.e., minimizing the $\mathcal{H}_2$ norm or 
$||\bm \sigma||_\infty$, the mean first hitting time can be effectively increased.
However, this is insufficient when compared with the one minimizing $||\bm P||_{\infty}$, which as shown is the most effective way to increase the mean first hitting time. This is because both the synchronous state determined in the deterministic system and the variance of phase differences determined in a stochastic system are considered in the objective of $||\bm P||_{\infty}$. In addition, it is found that the mean first hitting time decreases to 39s and 57.631s in the solution of 
the first two optimization problems respectively. In other words, maximizing the order parameter or the phase cohesiveness may decrease the synchronization stability. Hence, \emph{a larger order parameter or a higher level phase cohesiveness does not mean that the system is more robust against 
disturbances and it may not be wise to design the coupling strength of the network with disturbances so as to 
maximize these objectives. }
\par 
It is seen in Table \ref{Thevulnerablelines} for the solution of the 5 optimization problems that the most vulnerable edges
 which have the largest value of $r_k$  are $e_8, e_1,e_3, e_7, e_8$ respectively.  Clearly, these edges have been identified by the defined value $\widetilde{p}_k$.  Fig. \ref{objective1-result} shows the fluctuations of the phase differences around the values at 
the synchronous state at time 10-15s in the initial model and the 5 most vulnerable edges after designing the coupling strength with the 5 different objectives, respectively.  It is shown in Fig.\ref{objective1-result}(a-c) that 
the phase differences at the synchronous state, which are denoted by the dashed red lines, 
are effectively decreased by either maximizing the order parameter or the 
phase cohesiveness.  However, the variance of the phase difference is unexpectedly increased which leads to a high risk of the state hitting the boundary of the secure domain and a smaller mean first hitting time. This is also demonstrated 
by the data in Table \ref{design-coupling-strength}.  In contrast, by comparing 
the plots in Fig.\ref{objective1-result}(d-e) with the one in Fig.\ref{objective1-result}(a), it is found that the variance of the phase difference is greatly decreased by minimizing the $\mathcal{H}_2$ norm and $||\bm \sigma||_\infty$, which however does not 
effectively decrease the absolute value of the phase differences at the synchronous state. This further leads to a smaller mean first hitting time compared with the solution of the proposed optimization method as shown in Table \ref{design-coupling-strength}.
In particular, it is found that the fluctuations of the dynamics in Fig.\ref{objective1-result}(d-e) is much smaller than in Fig.\ref{objective1-result}(f), while the latter one have a longer mean first hitting time. \emph{This indicates that 
smaller fluctuations in the phase difference do not mean a stronger synchronization stability, where the 
expectation of the phase difference has to be considered. } 
\par 
Let us consider the design of the natural frequency by the 5 optimization frameworks.  
Table \ref{natural frequencies} shows the natural frequencies at nodes $1,2,3$ after solving the 5 optimization problems.
Table \ref{design-natural-frequency} shows the values of the objectives, 
the mean first hitting time and the values of $\mu_k, \sigma_k^2, \tilde{p}_k, r_k$ in the edge $e_k$ for $k=1,\cdots,m$. 
It is observed that minimizing 
the risk of the state hitting the boundary of the secure domain measured by $||\bm P||_{\infty}$ can effectively increase
the mean first hitting time.  When observing the order parameter $r$ and $||\bm \mu||_\infty$, it is found again that \emph{a larger order parameter or a smaller 
$||\bm \mu||_\infty$ does not mean a stronger synchronization stability}. Hence, it is demonstrated again that considering the variance of the phase differences only is
insufficient for increasing the synchronization stability. In the proposed 
optimization framework, because both the synchronous state that determined in a deterministic system and the fluctuations
of the phase differences in a stochastic system are considered, the synchronization stability can be effectively enhanced. 
In addition, it is demonstrated in Table \ref{vulnerableline-naturalfrequency} again that 
the most vulnerable edge can be effectively identified by the probability of the phase difference hitting the 
boundary of the secure domain. 
\par 
\begin{figure}[ht]
    \centering
	\includegraphics[width=0.34\textwidth]{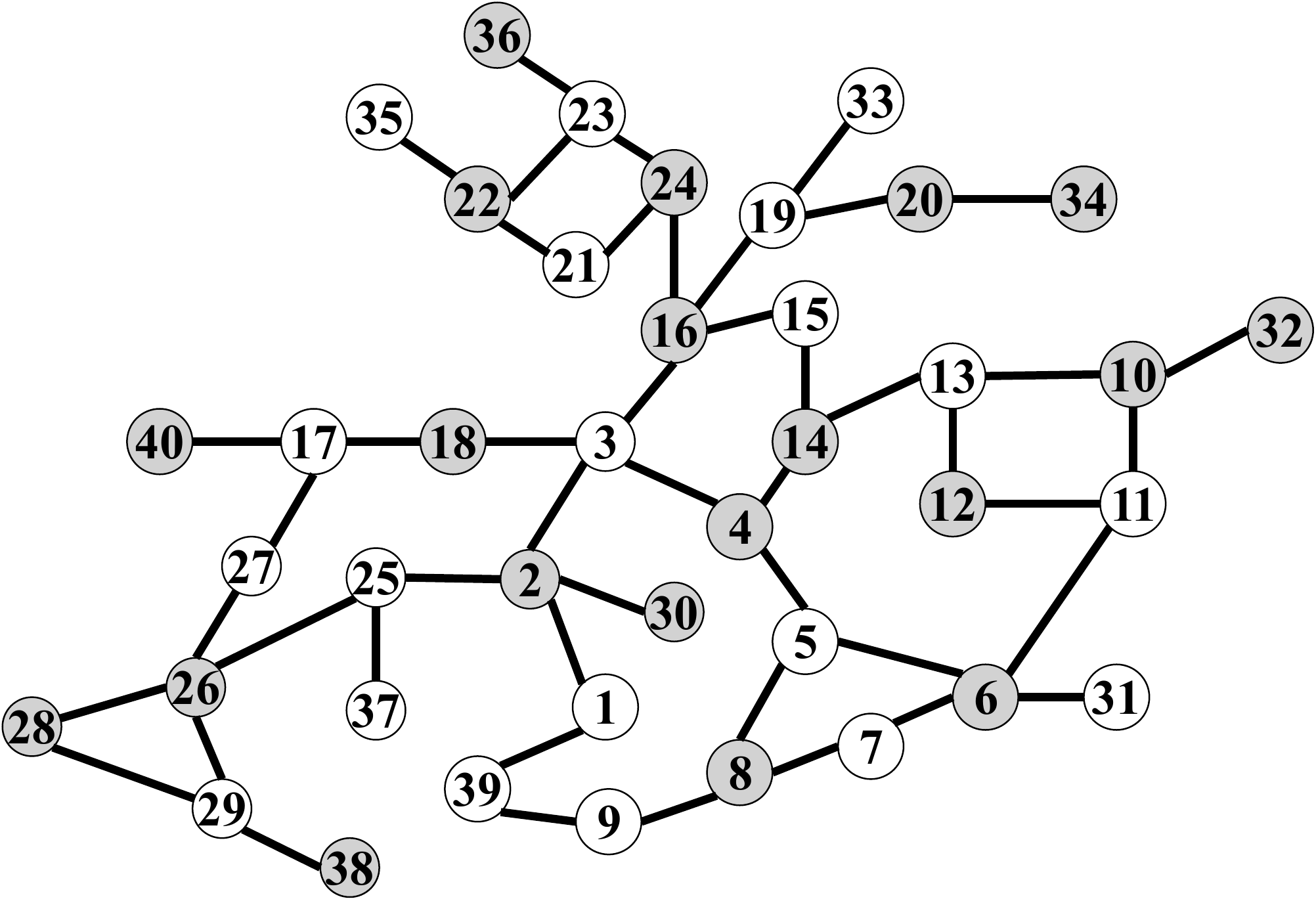}
    \caption{A network with 40 nodes and 47 edges. }\label{figure-net2}
\end{figure}

\begin{example}\label{example2}
Consider the network in Fig. \ref{figure-net2} with 40 nodes and 47 edges, which is generated randomly 
with the connecting probability between each pair of nodes being 0.06.  There are 20 grey nodes which are selected 
randomly and indexed by the even numbers.  
We set $T=10^5$, $\text{d}t=10^{-3}$, $N=10^5$, $b_i=0.95$ for all the nodes. 
We formulate an Initial model, in which we set $\omega_i=-3$ for the grey nodes and $\omega_i=3$ for the other nodes and $l_{ij}=10$ for all the edges. 
In the optimization problems for the design of the coupling strength, we set $\omega_i=-3$ for the grey nodes and $\omega_i=3$ for the other nodes, $W=470$ and $\underline{l}_{ij}=1$, 
$\overline{l}_{ij}=20$ for all the edges.  
In the optimization problems for the design of the natural frequency, we set 
$\underline{\omega}_i=-3$ and $\overline{\omega}_i=-3$ for the grey nodes and 
$\underline{\omega}_i=0$ and $\overline{\omega}_i=14$ for the other nodes and $l_{ij}=10$ for all the edges.
\end{example}

For the design of the coupling strength and the natural frequency, the values of the objective functions of the 
5 optimization problems are shown in Table \ref{solutionsfornetwork1} and Table \ref{solutionsfornetwork2} respectively. 
As in the results of the network in Example \ref{example1}, 
by maximizing 
the order parameter $r$ and the phase cohesiveness measured by $\|\mu\|_\infty$ for the design 
of the coupling strength,
the mean first hitting time decreases from about 85s to about 62s and to about 65s respectively. This indicates again 
that a larger order parameter or a higher level phase cohesiveness does not
means that the system is more robust against the disturbances. However, using the proposed optimization framework, the mean first hitting time 
increases from about 85s to about 4021s, which is much more effective than
minimizing the largest variance of the phase differences and the $\mathcal{H}_2$ norm. 
In addition, in the design 
of the natural frequency, the mean first hitting time increases from about 85s to about 494s by the proposed optimization framework. These findings demonstrates that 
the proposed optimization framework can effectively increase the synchronization stability of 
the complex system.

\section{Conclusion}\label{sec:conclusion}

In this paper, based on the theory of the invariant probability distribution of stochastic Gaussian processes, we have proposed a new metric for the synchronization stability of complex networks, that is 
the probability of the state being absent from a secure domain. By this metric, the most vulnerable edges
that may lead to desynchronization can be precisely identified. Using this metric
as the objective functions of optimization problems, either
the natural frequencies or the coupling strength can be assigned to improve the synchronization stability.
It is demonstrated in the case studies that by optimizing this metric, the mean first hitting time when the state of the system under stochastic disturbances hits the boundary of the secure domain can be effectively increased.
In contrast,  optimization of either the order parameter or the phase cohesiveness defined for a deterministic model may dramatically 
decrease the mean first hitting time and further decreases the synchronization stability. This indicates that it
is more practical to study the synchronization stability with the consideration 
of the strength of the disturbances as in the stochastic process. 
\par 
However, 
compared with the traditional methods for the synchronization stability analysis in the deterministic model, the strength of the disturbances has to be identified in the model of the stochastic processes and a 
matrix spectral decomposition is needed to compute the invariant probability 
distribution of the phase difference. In order to apply the proposed optimization framework to 
improve the synchronization stability of a large scale system in practice, efficient algorithms for the spectral decomposition and
for the optimization problems are important, which are the focus of the future research.


\appendix

\section{The optimization problems for the case study}\label{optimization-Case-study}
The order parameter of the couple phase oscillators is defined 
as $$r e^{\text{i}\phi}=\frac{1}{n}\sum_{j=1}^{n}e^{\text{i}\varphi_j},$$ where $\text{i}^2=-1$ and $\varphi_j$ 
is the phase at node $j$ and $r e^{\text{i}\phi}$ is the phase' centroid on the 
complex unit circle with the magnitude $r$ ranging from 0 to 1 \cite{kuramotobook}. 
In the case study, the order parameter is maximized by solving the following optimization problem \cite{Skardal2014},
\begin{equation}\label{orderoptimization1}
    \begin{aligned}
&\min_{l_{ij}\in\mathbb{R},(i,j)\in\mathcal{E}}r=1-||\bm\varphi||^2/n,\\
&\text{s.t}~~(\ref{totalcouplingstrength}),(\ref{opitmizaion-b}),\\
&~~~~~~\bm\varphi^*=\bm L^{\dagger}\bm\omega.
\end{aligned}
\end{equation}
where the matrix $\bm L^{\dagger}$ is defined in (\ref{linearstochastic}),
and the one for the design of the natural frequency is 
\begin{equation}\label{orderoptimization2}
    \begin{aligned}
&\min_{\omega_i\in\mathbb{R},i\in\mathcal{V}}r=1-||\bm\varphi^*||^2/n,\\
&\text{s.t}~~(\ref{constraintontotalfrequency0}),(\ref{constraintontotalfrequency}),\\
&~~~~~~\bm\varphi^*=\bm L^{\dagger}\bm\omega.
\end{aligned}
\end{equation}

The optimization problem for the design of the coupling strength with the objective 
of increasing the phase cohesiveness is 
\begin{equation}\label{cohesivenessoptimization1}
    \begin{aligned}
&\min_{l_{ij}\in\mathbb{R},(i,j)\in\mathcal{E}}||\bm y^*||_\infty,\\
&\text{s.t}~~\text{(\ref{synchronous state}),(\ref{expectation})},
(\ref{totalcouplingstrength}),(\ref{opitmizaion-b}),
\end{aligned}
\end{equation}
and the one for the design of the natural frequency with this objective is 
\begin{equation}\label{cohesivenessoptimization2}
    \begin{aligned}
&\min_{\omega_i\in\mathbb{R},i\in\mathcal{V}} ||\bm y^*||_\infty,\\
&\text{s.t}~~\text{(\ref{synchronous state}),(\ref{expectation})},(\ref{constraintontotalfrequency0}),
(\ref{constraintontotalfrequency}).
\end{aligned}
\end{equation}
\par

In section of case study,  the optimization problem for designing the 
coupling strength with the objective of minimizing the $\mathcal{H}_2$ norm follows, 
\begin{equation}\label{H2normoptimization1}
    \begin{aligned}
&\min_{l_{ij}\in\mathbb{R},(i,j)\in\mathcal{E}} \text{tr}(\bm Q_{\widehat{y}}),\\
&\text{s.t}~~\text{(\ref{synchronous state}),(\ref{spectral}),(\ref{Q_2solution})},(\ref{variance matrix}),
(\ref{totalcouplingstrength}),(\ref{opitmizaion-b})
\end{aligned}
\end{equation}
and the one to redistribute the natural frequency with this objective is 
\begin{equation}\label{H2normoptimization2}
    \begin{aligned}
&\min_{\omega_i\in\mathbb{R},i\in\mathcal{V}} \text{tr}(\bm Q_{\widehat{y}}),\\
&\text{s.t}~~\text{(\ref{synchronous state}),(\ref{spectral}),(\ref{Q_2solution})},(\ref{variance matrix}),(\ref{constraintontotalfrequency0}),
(\ref{constraintontotalfrequency})
\end{aligned}
\end{equation}
If the maximum of the variances of the phase differences in the edges is minimized, 
the objective function is replaced by $||\bm\sigma||_\infty$ in the above two optimization problems. 
\par 
\section{The invariant probability distribution and $\mathcal{H}_2$ norm}\label{InvariantDistribution}
Consider a linear time-invariant system, 
\begin{subequations}\label{Appendix:generalform}
 \begin{align}
  \dot{\bm x}&=\bm A\bm x+\bm B\bm w,\\
  \bm y&=\bm C\bm x, 
 \end{align}
\end{subequations}
where $\bm x\in\mathbb{R}^{n_x}$, $\bm A\in\mathbb{R}^{n_x\times n_x}$ is Hurwitz, $\bm B\in\mathbb{R}^{n_x\times n_w}$,  $\bm C\in\mathbb{R}^{n_y\times n_x}$, 
the input is denoted by $\bm w\in\mathbb{R}^{n_w}$ and the output of the system is denoted by $\bm y\in\mathbb{R}^{n_y}$. 
The squared $\mathcal{H}_2$ norm of the transfer matrix $\bm G$ of the mapping $(\bm A,\bm B,\bm C)$ from the input $\bm w$ to the output $\bm y$ is defined as 
\begin{subequations}\label{Appendix:H2norm}
 \begin{align}
  &||\bm G||^2_2=\text{tr}(\bm B^T\bm Q_o\bm B)=\text{tr}(\bm C\bm Q_c\bm C^T),\\
  &\bm Q_o\bm A+\bm A^T\bm Q_o+\bm C^T\bm C=\bm 0,\label{Lyapunov:equation}\\
  &\bm A\bm Q_c+\bm Q_c\bm A^T+\bm B\bm B^T=\bm 0,\label{Lyapunov:equation2}
 \end{align}
\end{subequations}
where $\text{tr}(\bm \cdot)$ denotes the trace of a matrix, $\bm Q_o,\bm Q_c\in\mathbb{R}^{n_x\times n_x}$ are the \emph{observability Grammian} of $(\bm C,\bm A)$ and \emph{controllability 
Grammian} of $(\bm A,\bm B)$ respectively \cite{H2norm_another_form,H2norm_book_toscano}. 
When the input $\bm w$ is modelled by Gaussian white noise,  the distribution of the state $\bm x$ and the output $\bm y$ are also Gaussian. 
Denote then for all $t \in T$,
$\bm x(t) \in G(\bm m_x(t), ~ \bm Q_{x}(t))$ with $\bm{Q}_{x}(t)\in \mathbb{R}^{n_x \times n_x}$ and 
$\bm y(t) \in G(\bm m_y(t), ~ \bm Q_{y}(t))$ with $\bm{Q}_{y}(t)\in \mathbb{R}^{n_y \times n_y}$.
Because the matrix $\bm A$ is Hurwitz, there exists an invariant probability distribution
of this linear stochastic system 
with the representation and properties
\begin{eqnarray*}
\mathbf{0}
    & = & \lim_{t \rightarrow \infty} ~ \mathbf{m}_x(t), ~
          \bm 0 = \lim_{t \rightarrow \infty} ~ \bm m_y(t), 
          \nonumber \\
        \bm Q_x 
    & = & \lim_{t \rightarrow \infty} ~ \bm Q_{x}(t), 
          \bm Q_y = \lim_{t \rightarrow \infty} ~ \bm Q_{y}(t), 
\end{eqnarray*}
where the variance matrices are
\begin{eqnarray*}
\bm Q_x
   & = & \int_0^{+\infty} \exp(\bm A t) 
          \bm B \bm B^{\top}
         \exp(\bm A^{\top} t) 
          \text{d}t,~~
          \bm Q_y=\bm C \bm Q_x \bm C^{\top}.
\end{eqnarray*}
Here $\bm Q_x$ is the unique solution of the Lyapunov matrix function (\ref{Lyapunov:equation2}). 
\par 

\par

\bibliography{sample}
\newpage

\begin{table*}[htbp]
  \begin{center}
      \caption{The coupling strength $l_{ij}$,  the expectations $\mu_k$ and the variances $\sigma_k^2$ of the phase differences, the value $\widetilde{p}_k$ defined in (\ref{vulnerbleLines}), 
      the value $r_k$ defined in (\ref{ratioRk}), the
      mean first hitting time $\overline{t}_e$ and the values of the objective functions in the initial model and in the solutions of the 5 optimization problems with respect to the design of the coupling strength the network in Fig. \ref{figure1}.}
      \label{design-coupling-strength}
      \scalebox{0.9}{
      \begin{tabular}{|c|c|c|c|c|c|c|c|c|c|c|c|c|c|c|c|}
          \hline
          &~&$e_1$&$e_2$&$e_3$&$e_4$&$e_5$&$e_6$&$e_7$&$e_8$&$r$&$||\bm \mu||_\infty$&$||\bm \sigma||_\infty$&$\mathcal{H}_2$&$||\bm P||_\infty$&$\overline{t}_e$\\
          \hline
          \multirow{5}*{Init. Model} ~ &~~$l_{ij}$~~&8.000&8.000&8.000&8.000&8.000&8.000&8.000&8.000&\multirow{5}*{0.9576}&\multirow{5}*{0.539}&\multirow{5}*{0.055}&\multirow{5}*{0.367}&\multirow{5}*{3.601e-6}&\multirow{5}*{118.460s}				
         \\	
         \cline{2-10}
          &$\mu_k$&0.133 & -0.248 & 0.539 & -0.291 & -0.176 & 0.467 & 0.514 & -0.133 &~&~&~&~&~&~\\
		      \cline{2-10}
		      ~ &$\sigma_k^2$&0.051 & 0.038 & 0.045 & 0.036 & 0.045 & 0.046 & 0.055 & 0.051 &~&~&~&~&~&~\\
		      \cline{2-10}
		      ~ &$\widetilde{p}_k$&2.473e-5 & 1.192e-6 & 0.129 & 1.284e-6 & 4.909e-6 & 0.035 & 0.836 & 2.473e-5 &~&~&~&~&~&~\\
		      \cline{2-10}
		      ~ &$r_k$&0 & 0 & 0.179 & 0 & 0 & 0.070 & 0.751 & 0 &~&~&~&~&~&~\\
		       \hline
          \multirow{5}*{Max. $r$} &$l_{ij}$&4.882 & 10.673 & 11.990 & 6.417 & 5.037 & 11.999 & 11.998 & 1.005 
          &\multirow{5}*{0.9805}&\multirow{5}*{0.407}&\multirow{5}*{0.147}&\multirow{5}*{0.481}&\multirow{5}*{2.806e-4}&\multirow{5}*{39.000s}\\	
		      \cline{2-10} 
		      ~ &$\mu_k$&0.051 & -0.107 & 0.350 & -0.242 & -0.129 & 0.371 & 0.407 & -0.249 &~&~&~&~&~&~\\	      
		      \cline{2-10}
		      ~ &$\sigma_k^2$&0.096 & 0.033 & 0.033 & 0.036 & 0.051 & 0.038 & 0.047 & 0.147 &~&~&~&~&~&~\\	  
		      \cline{2-10}
		      ~ &$\widetilde{p}_k$&0.002 & 1.974e-12 & 4.253e-8 & 4.321e-9 & 3.363e-7 & 1.042e-6 & 1.319e-4 & 0.998 &~&~&~&~&~&~\\	    
		      \cline{2-10}
		      ~ &$r_k$&0.017 & 0 & 0 & 0 & 0 & 0 & 0.013 & 0.970 &~&~&~&~&~&~\\	   
          \hline
          \multirow{5}*{Min. $||\bm \mu||_\infty$ } &$l_{ij}$&1.807 & 10.146 & 11.023 & 5.086 & 8.604 & 11.855 & 11.888 & 3.592 &\multirow{5}*{0.9740}&\multirow{5}*{0.402}&\multirow{5}*{0.136}&\multirow{5}*{0.474}&\multirow{5}*{9.637e-5}&\multirow{5}*{57.631s}\\
		      \cline{2-10}						 		 	 	 	 
		      ~&$\mu_k$&0.196 & -0.108 & 0.397 & -0.289 & -0.082 & 0.371 & 0.402 & -0.098 &~&~&~&~&~&~\\
		       \cline{2-10}
		      ~&$\sigma_k^2$&0.136 & 0.034 & 0.036 & 0.037 & 0.040 & 0.035 & 0.046 & 0.111 &~&~&~&~&~&~\\
		       \cline{2-10}
		      ~&$\widetilde{p}_k$&0.948 & 1.310e-11 & 2.822e-6 & 1.071e-7 & 3.953e-10 & 7.339e-7 & 2.298e-4 & 0.052 &~&~&~&~&~&~\\
		       \cline{2-10}
		      ~&$r_k$&0.870 & 0 & 0.001 & 0 & 0 & 0 & 0.011 & 0.118 &~&~&~&~&~&~\\
          \hline
          \multirow{5}*{Min. $||\bm \sigma||_\infty$} &$l_{ij}$&8.979 & 5.538 & 9.332 & 4.711 & 8.218 & 8.804 & 9.429 & 8.990 &\multirow{5}*{0.9625}&\multirow{5}*{0.533}&\multirow{5}*{0.048}&\multirow{5}*{0.364}&\multirow{5}*{4.955e-7}&\multirow{5}*{363.396s}\\ 		 
		      \cline{2-10}
		      ~&$\mu_k$&0.108 & -0.225 & 0.533 & -0.308 & -0.142 & 0.450 & 0.441 & -0.108 &~&~&~&~&~&~\\
		      \cline{2-10}
		      ~&$\sigma_k^2$&0.046 & 0.044 & 0.045 & 0.044 & 0.045 & 0.046 & 0.048 & 0.046 &~&~&~&~&~&~\\
		       \cline{2-10}
		      ~&$\widetilde{p}_k$&7.622e-6 & 1.124e-4 & 0.720 & 0.001 & 1.196e-5 & 0.111 & 0.167 & 7.444e-6 &~&~&~&~&~&~\\
		       \cline{2-10}
		      ~&$r_k$&0 & 0 & 0.661 & 0.003 & 0 & 0.129 & 0.207 & 0 &~&~&~&~&~&~\\
	          \hline
          \multirow{5}*{Min. $\mathcal{H}_2$} &$l_{ij}$&8.112 & 6.111 & 9.660 & 5.739 & 7.600 & 9.080 & 9.586 & 8.112 &\multirow{5}*{0.9652}&\multirow{5}*{0.496}&\multirow{5}*{0.050}&\multirow{5}*{0.362}&\multirow{5}*{1.124e-7}&\multirow{5}*{773.220s}\\
		      \cline{2-10}					
		      ~ &$\mu_k$&0.112 & -0.217 & 0.496 & -0.279 & -0.155 & 0.435 & 0.441 & -0.112 &~&~&~&~&~&~\\
		       \cline{2-10}
		      ~ &$\sigma_k^2$&0.050 & 0.042 & 0.042 & 0.040 & 0.046 & 0.044 & 0.048 & 0.050 &~&~&~&~&~&~\\
		       \cline{2-10}
		      ~ &$\widetilde{p}_k$&1.438e-4 & 7.541e-5 & 0.374 & 2.422e-4 & 8.899e-5 & 0.138 & 0.487 & 1.438e-4 &~&~&~&~&~&~\\
		       \cline{2-10}
		      ~ &$r_k$&0 & 0 & 0.383 & 0 & 0.001 & 0.158 & 0.458 & 0 &~&~&~&~&~&~\\
          \hline
          \multirow{5}*{Min. $||\bm P||_\infty$} &$l_{ij}$&7.489 & 4.881 & 11.713 & 3.972 & 7.489 & 11.035 & 11.719 & 5.701 &\multirow{5}*{0.9749}&\multirow{5}*{0.429}&\multirow{5}*{0.064}&\multirow{5}*{0.373}&\multirow{5}*{4.302e-9}&\multirow{5}*{3951.733s}\\ 		
          \cline{2-10}
          ~&$\mu_k$&0.091 & -0.167 & 0.429 & -0.261 & -0.120 & 0.381 & 0.378 & -0.120 &~&~&~&~&~&\\ 
          \cline{2-10}
          ~ &$\sigma_k^2$&0.054 &	0.046 &	0.038 &	0.044 &	0.046 &	0.039 &	0.041 &	0.064 &~&~&~&~&~&~\\
		       \cline{2-10}
		      ~ &$\widetilde{p}_k$&0.011 & 0.003 & 0.229 & 0.023 & 6.564e-4 & 0.083 & 0.228 & 0.422 &~&~&~&~&~&~\\
		       \cline{2-10}
		      ~ &$r_k$&0.009 &	0.003 &	0.282 &	0.030 &	0.001 &	0.090 &	0.240 &	0.310 &~&~&~&~&~&\\
          \hline
      \end{tabular}
      }				
  \end{center}
\end{table*}

\begin{table*}[htbp]
  \begin{center}
      \caption{The most vulnerable edges identified by 4 metrics in the initial model and in the solution of the 5
optimization problems with respect to the design of the coupling strength for the network in Fig. \ref{figure1}. }
      \label{Thevulnerablelines}
      \begin{tabular}{|c|c|c|c|c|}
          \hline
          ~&by $\mu_k$&by $\sigma_k$&by $p_k$&by $r_k$\\
          \hline
          Init. Model&$e_3$&$e_7$&$e_7$&$e_7$\\
          \hline
          Max. $r$ &$e_7$&$e_8$&$e_8$&$e_8$\\
          \hline
          Min. $||\bm \mu||_\infty$ &$e_7$&$e_1$&$e_1$&$e_1$\\
          \hline
          Min. $||\bm \sigma||_\infty$&$e_3$&$e_7$&$e_3$&$e_3$\\
          \hline
          Min. $\mathcal{H}_2$&$e_3$&$e_1$&$e_7$&$e_7$\\
          \hline
          Min. $||\bm P||_\infty$&$e_3$&$e_8$&$e_8$&$e_8$\\
          \hline
      \end{tabular}
  \end{center}
\end{table*}

\begin{table*}[htbp]
  \begin{center}
      \caption{The natural frequencies at  node $i=1,\cdots, 3$  in the initial model and in the solutions of the 5 optimization problems 
      with respect to the design of the natural frequency for the network in Fig. \ref{figure1}. }
      \label{natural frequencies}
      \begin{tabular}{|c|c|c|c|}
          \hline
          &$\omega_1$&$\omega_2$&$\omega_3$\\
          \hline
          Init. Model&5.000& 5.000&5.000\\
          \hline
          Max. $r$ &1.128 & 0.487 & 13.385 \\
          \hline
          Min. $||\bm \mu||_\infty$&5.034 & 3.691 & 6.275 \\
          \hline
          Min. $||\bm \sigma||_\infty$&2.621 & 2.720 & 9.660 \\
          \hline
          Min. $\mathcal{H}_2$&2.503 & 2.977 & 9.520 \\
          \hline
          Min. $||\bm P||_\infty$&1.765 & 6.395 & 6.840 \\
          \hline
      \end{tabular}
  \end{center}
\end{table*}

\begin{table*}[htbp]
  \begin{center}
      \caption{The expectations $\mu_k$ and the variances $\sigma_k^2$ of the phase differences, the value $\widetilde{p}_k$ defined in (\ref{vulnerbleLines}), 
      the value $r_k$ defined in (\ref{ratioRk}), the
      mean first hitting time $\overline{t}_e$ and the values of the objective functions in the initial model and in the solutions of the 5 optimization problems with respect to the design of the natural frequency for the network in Fig. \ref{figure1}.}
      \label{design-natural-frequency}
      \scalebox{0.9}{
      \begin{tabular}{|c|c|c|c|c|c|c|c|c|c|c|c|c|c|c|c|}
          \hline
          &~&$e_1$&$e_2$&$e_3$&$e_4$&$e_5$&$e_6$&$e_7$&$e_8$&$r$&$||\bm \mu||_\infty$&$||\bm \sigma||_\infty$&$\mathcal{H}_2$&$||\bm P||_\infty$&$\overline{t}_e$\\
          \hline
          \multirow{5}*{Init. Model} ~&$\mu_k$&0.133 & -0.248 & 0.539 & -0.291 & -0.176 & 0.467 & 0.514 & -0.133 &\multirow{5}*{0.9576}&\multirow{5}*{0.539}&\multirow{5}*{0.055}&\multirow{5}*{0.367}&\multirow{5}*{3.601e-6}&\multirow{5}*{118.460s}		\\
		      \cline{2-10}
          ~ &$\sigma_k^2$&0.051 & 0.038 & 0.045 & 0.036 & 0.045 & 0.046 & 0.055 & 0.051 &~&~&~&~&~&~\\
		      \cline{2-10}
		      ~ &$\widetilde{p}_k$&2.473e-5 & 1.192e-6 & 0.129 & 1.284e-6 & 4.909e-6 & 0.035 & 0.836 & 2.473e-5 &~&~&~&~&~&~\\
		      \cline{2-10}
		      ~ &$r_k$&0 & 0 & 0.179 & 0 & 0 & 0.070 & 0.751 & 0 &~&~&~&~&~&~\\
		       \hline
          \multirow{5}*{Max. $r$} ~ &$\mu_k$&-0.042 & 0.231 & 0.251 & -0.482 & -0.087 & 0.569 & 0.184 & -0.458 &\multirow{5}*{0.9819}&\multirow{5}*{0.569}&\multirow{5}*{0.054}&\multirow{5}*{0.365}&\multirow{5}*{2.464e-6}&\multirow{5}*{151.223s}\\	      
		      \cline{2-10} 
		      ~ &$\sigma_k^2$&0.051 & 0.037 & 0.042 & 0.036 & 0.045 & 0.048 & 0.051 & 0.054 &~&~&~&~&~&~\\	  
		      \cline{2-10}
		      ~ &$\widetilde{p}_k$&1.829e-6 & 5.459e-7 & 2.103e-5 & 0.002 & 4.173e-7 & 0.738 & 1.359e-4 & 0.260 &~&~&~&~&~&~\\	    
		      \cline{2-10}
		      ~ &$r_k$&0 & 0 & 0 & 0.007 & 0 & 0.717 & 0 & 0.276 &~&~&~&~&~&~\\	   
          \hline
          \multirow{5}*{Min. $||\bm \mu||_\infty$ }~&$\mu_k$&0.164 & -0.160 & 0.484 & -0.324 & -0.160 & 0.484 & 0.484 & -0.160 &\multirow{5}*{0.9623}&\multirow{5}*{0.484}&\multirow{5}*{0.055}&\multirow{5}*{0.366}&\multirow{5}*{1.722e-6}&\multirow{5}*{203.074s}\\  
		       \cline{2-10} 	  
		      ~&$\sigma_k^2$&0.051 & 0.037 & 0.044 & 0.035 & 0.045 & 0.047 & 0.055 & 0.051 &~&~&~&~&~&~\\
		       \cline{2-10}
		      ~&$\widetilde{p}_k$&1.255e-4 & 6.565e-8 & 0.055 & 8.819e-6 & 6.133e-6 & 0.118 & 0.827 & 1.102e-4 &~&~&~&~&~&~\\
		       \cline{2-10}
		      ~&$r_k$&0 & 0 & 0.096 & 0 & 0.001 & 0.157 & 0.745 & 0.001 &~&~&~&~&~&~\\
          \hline
          \multirow{5}*{Min $||\bm \sigma||_\infty$}&$\mu_k$&0.015 & 0.015 & 0.378 & -0.393 & -0.128 & 0.521 & 0.318 & -0.318 &\multirow{5}*{0.9778}&\multirow{5}*{0.521}&\multirow{5}*{0.052}&\multirow{5}*{0.362}&\multirow{5}*{6.740e-7}&\multirow{5}*{449.385s}\\
		      \cline{2-10}
		      ~&$\sigma_k^2$&0.051 & 0.037 & 0.043 & 0.036 & 0.045 & 0.047 & 0.052 & 0.052 &~&~&~&~&~&~\\
		       \cline{2-10}
		      ~&$\widetilde{p}_k$&4.301e-6 & 4.628e-10 & 0.006 & 3.104e-4 & 6.512e-6 & 0.937 & 0.029 & 0.028 &~&~&~&~&~&~\\
		       \cline{2-10}
		      ~&$r_k$&0 & 0 & 0.009 & 0.001 & 0 & 0.858 & 0.065 & 0.067 &~&~&~&~&~&~\\
	        \hline
          \multirow{5}*{Min. $\mathcal{H}_2$}
		      ~ &$\mu_k$&0.001 & 0.003 & 0.386 & -0.388 & -0.130 & 0.518 & 0.317 & -0.319 &\multirow{5}*{0.9777}&\multirow{5}*{0.518}&\multirow{5}*{0.052}&\multirow{5}*{0.362}&\multirow{5}*{6.289e-7}&\multirow{5}*{469.604s}\\
          \cline{2-10}
		      ~ &$\sigma_k^2$&0.051 & 0.037 & 0.043 & 0.036 & 0.045 & 0.047 & 0.052 & 0.052 &~&~&~&~&~&~\\
		       \cline{2-10}
		      ~ &$\widetilde{p}_k$&4.118e-6 & 3.287e-10 & 0.008 & 2.781e-4 & 7.420e-6 & 0.931 & 0.030 & 0.031 &~&~&~&~&~&~\\
		       \cline{2-10}
		      ~ &$r_k$&0 & 0 & 0.025 & 0 & 0 & 0.872 & 0.053 & 0.050 &~&~&~&~&~&~\\
          \hline
          \multirow{5}*{Min. $||\bm P||_\infty$}~&$\mu_k$&-0.125 & -0.193 & 0.505 & -0.312 & -0.166 & 0.478 & 0.352 & -0.284 &\multirow{5}*{0.9720}&\multirow{5}*{0.505}&\multirow{5}*{0.053}&\multirow{5}*{0.364}&\multirow{5}*{2.052e-7}&\multirow{5}*{550.514s}\\ 
          \cline{2-10}
          ~ &$\sigma_k^2$&0.051 & 0.037 & 0.044 & 0.035 & 0.045 & 0.047 & 0.053 & 0.052 &~&~&~&~&~&~\\
		       \cline{2-10}
		      ~ &$\widetilde{p}_k$&1.528e-4 & 1.074e-6 & 0.433 & 2.441e-5 & 3.255e-5 & 0.433 & 0.116 & 0.017 &~&~&~&~&~&~\\
		       \cline{2-10}
		      ~ &$r_k$&0 & 0 & 0.418 & 0.001 & 0 & 0.426 & 0.127 & 0.028 &~&~&~&~&~&\\
          \hline
      \end{tabular}
      }
  \end{center}
\end{table*}

\begin{table*}[htbp]
  \begin{center}
      \caption{The most vulnerable edges identified by 4 metrics in the initial model and in the solutions of the 5
optimization problems with respect to the design of the natural frequency for the network in Fig. \ref{figure1}}
      \label{vulnerableline-naturalfrequency}
      \begin{tabular}{|c|c|c|c|c|}
          \hline
          ~&by $u_k$&by $\sigma_k$ &by $p_k$&by $r_k$\\
          \hline
          Init. Model&$e_3$&$e_7$&$e_7$&$e_7$\\
          \hline
          Max. $r$ &$e_6$&$e_8$&$e_6$&$e_6$\\
          \hline
          Min. $||\bm \mu||_\infty$ &$e_7$&$e_7$&$e_7$&$e_7$\\
          \hline
          Min. $||\bm \sigma||_\infty$&$e_6$&$e_7$&$e_6$&$e_6$\\
          \hline
           Min. $\mathcal{H}_2$&$e_6$&$e_8$&$e_6$&$e_6$\\
          \hline
          Min. $||\bm P||_\infty$&$e_3$&$e_7$&$e_3$&$e_3$\\
          \hline
      \end{tabular}
  \end{center}
\end{table*} 
\par 
\begin{table*}[htbp]
    \begin{center}
        \caption{The
      mean first hitting time $\overline{t}_e$ and the values of the objective functions in the initial model and in the solutions of the 5 optimization problems with respect to the design of the coupling strength for the network in Fig. \ref{figure-net2}}
        \label{solutionsfornetwork1}
         \scalebox{0.9}{
        \begin{tabular}{|c|c|c|c|c|c|c|}
            \hline
            ~&$r$&$||\bm \mu||_\infty$&$||\bm \sigma||_\infty$&$\mathcal{H}_2$&$||\bm P||_\infty$&$\overline{t}_e$\\
            \hline
            {Init. Model} ~ &0.9148	&0.644	&	0.056	&	1.824	&	4.723e-5	&	84.572s			\\	
           \hline
            {Max. $r$ } &0.9685	&0.338		&0.178	&	2.683&		1.989e-4	&	62.164s\\ 	
            \hline 
            {Min. $||\bm \mu||_\infty$ } &0.9361	&0.330	&	0.059	&	2.188	&	1.679e-4	&	65.439s\\
            \hline
            {Min. $||\bm \sigma||_\infty$} &0.9259	&0.489		&0.039	&	1.811	&	1.826e-8	&	2022.807s\\ 	
            \hline
            {Min. $\mathcal{H}_2$} &0.9276	&0.491	&	0.041	&	1.802		&3.601e-8		&1738.061s\\
            \hline
            {Min. $||\bm P||_\infty$} &0.9238	&0.450		&0.054		&1.824		&1.372e-9	&	4021.587s
            \\ 		
            
            \hline
        \end{tabular}
        }				
    \end{center}
  \end{table*} 
  \begin{table*}[htbp]
    \begin{center}
             \caption{The
      mean first hitting time $\overline{t}_e$ and the values of the objective functions in the initial model and in the solutions of the 5 optimization problems with respect to the design of the natural frequency for the network in Fig. \ref{figure-net2}}\label{solutionsfornetwork2}
         \scalebox{0.9}{
        \begin{tabular}{|c|c|c|c|c|c|c|}
            \hline
            ~&$r$&$||\bm \mu||_\infty$&$||\bm \sigma||_\infty$&$\mathcal{H}_2$&$||\bm P||_\infty$&$\overline{t}_e$\\
            \hline
            {Init. Model} ~ &0.9148	&0.644	&	0.056	&	1.824	&	4.723e-5	&	84.572s\\	
           \hline
           
            {Max. $r$ } &0.9735	&0.653	&	0.056	&	1.793	&	2.435e-5	&	130.249s\\ 	
            \hline 
            {Min. $||\bm \mu||_\infty$ } &0.9329&	0.536	&	0.056	&	1.795	&	1.364e-5	&	163.661s\\
            \hline
            {Min. $||\bm \sigma||_\infty$} &0.9445	&0.643	&	0.051	&	1.784	&	5.969e-6	&	287.234s\\ 	
            \hline
            {Min. $\mathcal{H}_2$} &0.9446	&0.643	&	0.051	&	1.782	&	5.820e-6	&	294.689s\\
            \hline
            {Min. $||\bm P||_\infty$} &0.9412	&0.640	&	0.053	&	1.795	&	1.543e-6	&	493.585s\\ 		
            
            \hline
        \end{tabular}
        }				
    \end{center}
  \end{table*}

\end{document}